\documentclass[manuscript,screen,nonacm]{acmart}

\setcopyright{none}
\AtBeginDocument{%
  }

\usepackage{amsmath}
\usepackage{graphicx}

\usepackage[most]{tcolorbox}

\usepackage{booktabs}

\usepackage{amssymb}

\usepackage{longtable}
\usepackage{booktabs}
\usepackage{array}
\usepackage{hyperref}

\usepackage{array}
\usepackage{colortbl}

\usepackage[table]{xcolor}
\usepackage{array}

\usepackage{tabularx,booktabs}

\usepackage{array,booktabs}

\usepackage{pifont} 
\newcolumntype{C}[1]{>{\centering\arraybackslash}m{#1}}
\newcolumntype{B}{>{\centering\arraybackslash\columncolor{blue!10}}m{0.7cm}} 

\begin{document}


\title[SVM: SLR]{An Industry Perspective on Software Vulnerability Management: A Systematic Review of Approaches, Tools, Practices, and Challenges}

\author{M. Mehdi Kholoosi}
\email{mehdi.kholoosi@adelaide.edu.au}
\orcid{0000-0002-5712-1378}


\affiliation{%
  \institution{Adelaide University}
  \city{Adelaide}
  \state{South Australia}
  \country{Australia}
}

\author{Triet Huynh Minh Le}
\email{triet.h.le@adelaide.edu.au}
\affiliation{%
 \institution{Adelaide University}
 \city{Adelaide}
 \state{South Australia}
 \country{Australia}}

\author{M. Ali Babar}
\email{ali.babar@adelaide.edu.au}

\affiliation{%
  \institution{Adelaide University}
  \city{Adelaide}
  \state{South Australia}
  \country{Australia}
}
\affiliation{%
  \institution{Elevexai Systems}
  \city{Adelaide}
  \state{South Australia}
  \country{Australia}
}


\renewcommand{\shortauthors}{Kholoosi et al.}

\begin{abstract}
Software vulnerability management (SVM) refers to the coordinated set of activities used to detect, assess, repair, and disclose software vulnerabilities (SVs) across a product or ecosystem. As software systems grow in scale and dependency depth, organisations must balance timely security action with limited expertise, evolving tooling, and delivery pressure. A clearer understanding of what practitioner-centred evidence says about existing SVM solutions and real-world obstacles can help align research, tool building, and organisational investment. In this practitioner-centred systematic literature review, we synthesise 57 peer-reviewed empirical studies published between 2010 and 2025, including only work that collects primary data directly from industry practitioners. We analyse these studies to characterise what types of SVM solutions (approaches, tools, and practices) have been proposed and evaluated with practitioners, and what challenges practitioners report when adopting and using such solutions in real-world settings. We find that practice-oriented contributions such as guidelines, recommendations, lessons learned, and shared experiences constitute most of the evidence base, while fewer studies evaluate concrete approaches and tools in industrial environments. The solution landscape and evaluation focus are weighted toward SV detection and repair, with comparatively limited practitioner-centred coverage of SV assessment and disclosure. Across the corpus, the most persistent challenges stem from how SVM work is organised and embedded: limited security expertise and time, uncertainty about who is responsible for SV-related decisions and work, and processes that are scattered or poorly coordinated, together with tools that are hard to understand, configure, or integrate into existing workflows. These findings highlight the need for more integrated socio-technical interventions, stronger coverage of under-studied SVM tasks, and evaluations that prioritise sustained workflow impact alongside technical performance.
\end{abstract}




\maketitle

\section{Introduction}
\label{Sec:Intro}

Software systems are routinely exposed to security weaknesses that can be discovered, weaponised, and exploited at scale. Software vulnerabilities (SVs) are commonly understood as weaknesses in the computational logic of software (e.g., code) that adversaries can exploit to compromise confidentiality, integrity, or availability \cite{allen2008software,nist_nvd}. When exploited, such SVs can cause severe operational and financial damage. For example, in May 2025, a rounding bug in a shared library used by the Cetus decentralised finance protocol was exploited to drain an estimated \$223 million in assets within minutes, illustrating how a single SV in critical infrastructure can trigger large-scale financial losses \cite{merklescience2025cetus}. The number and complexity of SVs continue to increase \cite{kholoosi-timing,le2022survey}, and many remain unpatched for prolonged periods, with an empirical study reporting that a substantial fraction of known SVs take more than a year to remediate \cite{iannone2022secret}. In this landscape, defending modern software is not only a matter of building detection algorithms or scoring models, but of managing SVs as part of everyday engineering and operational work. Organisations must detect weaknesses, assess their relevance and risk in context, repair affected components, and communicate about discovered issues and their remediation status to stakeholders over time. Following prior work, we refer to this coordinated set of activities as Software Vulnerability Management (SVM) \cite{dowd2006art,le2022survey} and treat it as an iterative life cycle with four phases: \emph{detection} (identifying potential SVs), \emph{assessment} (analysing their impact and risk), \emph{repair} (implementing changes that eliminate the SVs or reduce their exploitability), and \emph{disclosure} (communicating about SVs and their remediation status to affected stakeholders) \cite{shahzad2012large,shahzad2019large}. We use this four-phase SVM life cycle as the organising frame for the remainder of the paper.

Recent years have seen a surge of technical advances in SVM-related capabilities. Data-driven and Deep Learning (DL) approaches for SV detection and prediction promise to uncover patterns in code and artefacts \cite{ghaffarian2017software,croft2022data,sonnekalb2022deep,shiri2024systematic}. Automated assessment techniques apply Machine Learning (ML) and Natural Language Processing (NLP) to derive severity scores, exploitability estimates, and SV types from structured and unstructured data \cite{le2022survey,elder2024survey}. In parallel, repair techniques frame SV fixing as a code-transformation problem and increasingly explore Transformer-based models and Large Language Models (LLMs) to propose patches or repair candidates \cite{zhou2025large,fu2025ai,kholoosi2024qualitative}. These capabilities are increasingly explored in research prototypes and industrial experiments that target the integration of these capabilities into developer tooling, CI/CD pipelines, and DevSecOps practices, with a growing focus on automation and workflow support \cite{fu2025ai,Steenhoek2025}.

Alongside these technical developments, a growing body of secondary studies has mapped approaches, datasets, and evaluation setups for individual parts of the SVM life cycle. Existing surveys and Systematic Literature Reviews (SLRs) examine SV detection and prediction \cite{ghaffarian2017software,croft2022data,sonnekalb2022deep,bassi2023systematic,kalouptsoglou2023software,senanayake2023android,shiri2024systematic,shimmi2025ai,wang2025review}, data-driven SV assessment and exploitability estimation \cite{le2022survey,elder2024survey}, and security patch management and repair in both technical and socio-technical settings \cite{dissanayake2022software,lin2024vulnerabilities}. Many of these reviews focus on learning-based or AI-driven techniques for these tasks, and recent work specifically examines the role of AI and LLMs in security-oriented DevSecOps workflows \cite{fu2025ai,zhou2025large}. As discussed in Section~\ref{Sec:Background}, these reviews provide thorough coverage of technical approaches, benchmarks, and research trends for specific SVM tasks.

However, comparatively little is known about how SVM solutions behave once they leave the lab and are embedded in real organisations. Without practitioner-centred evidence, it is difficult to tell whether an SV detection tool that performs well on benchmarks actually ends up disabled because it floods developers with unactionable alerts, or whether a patch-management workflow that looks sound on paper is routinely bypassed to avoid downtime and missed release deadlines. Existing reviews on ML- and DL-based SV detection and prediction primarily synthesise algorithms, models, datasets, and evaluation metrics, and typically rely on secondary artefacts such as code repositories, SV databases, and benchmarks as their main evidence base \cite{ghaffarian2017software,sonnekalb2022deep,shiri2024systematic,wang2025review}. Reviews that touch on patch management and SV assessment provide important insights into remediation workflows and scoring approaches, but they remain scoped to specific activities rather than offering a practitioner-centred synthesis across the SVM life cycle \cite{dissanayake2022software,le2022survey}. In these existing reviews, issues such as how responsibility for SV-related tasks (for example, prioritisation and patching) is allocated and coordinated across teams, how SVM tools and processes are embedded into development and operations workflows, and what obstacles practitioners report when managing SVs over time receive only limited attention. As a result, there is still a gap between the rapidly evolving technical landscape of SVM research and the socio-technical realities faced by practitioners who must integrate SVM activities into product roadmaps, release trains, and other operational constraints.

This paper addresses that gap through a practitioner-centred SLR of SVM. We focus exclusively on empirical studies that collect primary data from industry practitioners (e.g., developers, security engineers) and that target at least one phase of the SVM life cycle. Using a protocol-based search and screening process, we identify 57 primary studies published in top-ranked software engineering and security venues. From these, we extract (i) approaches and tools that have been introduced or evaluated with practitioners and (ii) practices that describe how SVM work is carried out in real-world settings. We then conduct a thematic analysis of practitioner-reported challenges to adopting and using SVM solutions in real-world contexts.

Our review makes three main contributions:

\begin{itemize}
    \item It maps the \emph{practitioner-centred solution space} of SVM by considering only approaches, tools, and practices that have been applied or evaluated with industry practitioners. We distinguish between technical approaches and tools on the one hand and practice-oriented contributions (e.g., guidelines, recommendations, lessons learned, and best practices) on the other, and relate them to the four phases of the SVM life cycle, complementing existing reviews that primarily catalogue technical solutions and benchmarks.

    \item It synthesises \emph{practitioner-reported challenges} into a set of data-driven themes that characterise how organisational conditions, tool adoption and integration, information and risk assessment, system architecture and dependencies, and constraints on detection, scanning, and review shape SVM work in practice.

    \item It relates the identified solutions and challenges to highlight alignment and gaps across the SVM life cycle, and derives implications and future research directions for SVM research, practice, and tool building, with particular attention to socio-technical interventions beyond individual tools.
\end{itemize}

The remainder of the paper is structured as follows. Section~\ref{Sec:Background} introduces SVM and situates our work with respect to existing SVM-related reviews. Section~\ref{Sec:Method} describes the methodology, including the protocol, research questions, search strategy, selection criteria, and data analysis procedures. Section~\ref{Sec:Result} presents the results for the solution space (RQ1) and practitioner-reported challenges (RQ2). Section~\ref{sec:discussion} discusses implications, positions our findings relative to prior work, and outlines future research directions. Section~\ref{Sec:threats-validity} reflects on threats to validity, and Section~\ref{sec:conclusion} concludes the paper.

\section{Background and related work}
\label{Sec:Background}


\subsection{Software vulnerability management and related reviews}
Software vulnerability management (SVM) is commonly understood as a life cycle spanning four interrelated phases: \textit{detection}, \textit{assessment}, \textit{repair}, and \textit{disclosure}. We use this four-phase structure to position prior secondary studies and to organise the practitioner-centred synthesis that follows. Table~\ref{tab:SLR-comparison} provides an overview of existing reviews related to SVM, including their phase coverage, contribution, and the extent to which they address practitioner-facing concerns.

\begin{table}[htbp]
\centering
\caption{Comparison of contributions between our study and the existing related reviews.
SVM task abbreviations: D = Detection, A = Assessment, R = Repair, Di = Disclosure.}
\label{tab:SLR-comparison}
\resizebox{\textwidth}{!}{%
\begin{tabular}{|C{3.7cm}|C{6.4cm}|C{2.2cm}|C{1.1cm}|C{3.1cm}|B|B|B|B|}
\hline
\textbf{Study} & \textbf{Study contribution} & \textbf{Time} & \textbf{\#P} & \textbf{Practitioner-Centered Lens} &
\multicolumn{4}{>{\columncolor{blue!10}}c|}{\textbf{SVM Task Focus}} \\
\cline{6-9}
 &  &  &  &  & \textbf{D} & \textbf{A} & \textbf{R} & \textbf{Di.} \\
\hline
Ghaffarian and Shahriari 2017 \cite{ghaffarian2017software} & Survey of ML/data-mining approaches for vulnerability analysis and discovery. & 2007--2016 & -- & -- & \checkmark & -- & -- & -- \\
\hline
Dissanayake et al.\ 2022 \cite{dissanayake2022software} & SLR of security patch management challenges, approaches, and tools. & 2002--2020 & 72 & -- & -- & -- & \checkmark & -- \\
\hline
Le et al.\ 2022 \cite{le2022survey} & SLR on data-driven vulnerability assessment and prioritisation. & $\leq$2021 & 84 & -- & -- & \checkmark & -- & -- \\
\hline
Croft et al.\ 2022 \cite{croft2022data} & SLR of data preparation challenges for vulnerability prediction. & $\leq$2021 & 61 & -- & \checkmark & -- & -- & -- \\
\hline
Sonnekalb et al.\ 2022 \cite{sonnekalb2022deep} & SLR on deep program code security analysis approaches and tools. & 2017--2021 & 32 & -- & \checkmark & -- & -- & -- \\
\hline
Bassi and Singh 2023 \cite{bassi2023systematic} & SLR of software vulnerability prediction models using ML. & 2000--2022 & 77 & -- & \checkmark & -- & -- & -- \\
\hline
Kalouptsoglou et al.\ 2023 \cite{kalouptsoglou2023software} & Mapping study of research trends, datasets, and methods in vulnerability prediction. & $\leq$2022 & 180 & -- & \checkmark & -- & -- & -- \\
\hline
Senanayake et al.\ 2023 \cite{senanayake2023android} & SLR on Android source code vulnerability detection. & 2016--2022 & 118 & -- & \checkmark & -- & -- & -- \\
\hline
Lin et al.\ 2024 \cite{lin2024vulnerabilities} & Survey of vulnerability and patch detection methods in OSS. & 2014--2023 & 127 & -- & \checkmark & -- & \checkmark & -- \\
\hline
Elder et al.\ 2024 \cite{elder2024survey} & Survey on methods for exploitability assessment. & 2008--2022 & 76 & -- & -- & \checkmark & -- & -- \\
\hline
Harzevili et al.\ 2024 \cite{shiri2024systematic} & SLR of ML-based automated software vulnerability detection. & 2011--2024 & 138 & -- & \checkmark & -- & -- & -- \\
\hline
Zhou et al.\ 2025 \cite{zhou2025large} & SLR on LLMs for vulnerability detection and repair. & 2018--2024 & 58 & -- & \checkmark & -- & \checkmark & -- \\
\hline
Fu et al.\ 2025 \cite{fu2025ai} & Review of AI-driven security in DevSecOps. & 2017--2023 & 99 & -- & \checkmark & \checkmark & \checkmark & -- \\
\hline
Shimmi et al.\ 2025 \cite{shimmi2025ai} & SLR of AI-based vulnerability detection. & 2018--2023 & 98 & -- & \checkmark & -- & -- & -- \\
\hline
Wang et al.\ 2025 \cite{wang2025review} & SLR of learning-based smart contract vulnerability detection. & 2018--2024 & 61 & -- & \checkmark & -- & -- & -- \\
\hline
\rowcolor{gray!15}
\textbf{This study} & \textbf{Practitioner-centered SLR of Software Vulnerability Management.} & \textbf{$\leq$2025} & \textbf{57} & \checkmark & \checkmark & \checkmark & \checkmark & \checkmark \\
\hline
\end{tabular}%
}
\vspace{4pt}
\begin{minipage}{\linewidth}
\footnotesize\emph{Note.} A tick indicates an \textit{explicit} task focus in the corresponding review; incidental mentions are left blank. 
\end{minipage}
\end{table}

\subsubsection{Detection}
SV detection concerns the identification of security weaknesses in software systems before they can be exploited, with the aim of uncovering flaws that may compromise confidentiality, integrity, or availability \cite{le2022survey}. It can operate across levels of analysis, from source code and binaries to runtime behaviour, and may rely on automated or manual techniques such as symbolic execution, fuzz testing, taint analysis, and binary analysis \cite{lin2024vulnerabilities,fu2025ai,shiri2024systematic}. Detection approaches may be static, dynamic, or hybrid depending on whether the software is inspected without execution, observed at runtime, or assessed through a combination of both \cite{shiri2024systematic,ghaffarian2017software}. Prior studies on this task are dominated by work on AI-based detection, together with surveys of data preparation and domain-specific detection settings.

Ghaffarian and Shahriari \cite{ghaffarian2017software} organise early ML-based SV detection literature into four categories: prediction models based on software metrics, anomaly detection, vulnerable code pattern recognition, and miscellaneous techniques. Their scheme also separates approaches that analyse program syntax and semantics from those that do not, and highlights a barrier: the absence of standard benchmarking datasets and consistent labelling, which makes cross-paper comparison unreliable. Building on this foundation, Sonnekalb et al. \cite{sonnekalb2022deep} survey 32 studies on DL for code security and catalogue the representations used for code, from source or object code at different granularities to encodings such as one-hot vectors, word embeddings, and graph structures. They also summarise the reported network topologies, with recurrent models most common alongside CNNs and hybrid designs, and point to promising directions such as stronger code embeddings, broader use of graph representations, and more systematic evaluation of Transformer-based approaches. Harzevili et al. \cite{shiri2024systematic} take an end-to-end view across 138 studies, covering data collection, preprocessing, input representations, feature extraction, model choice, and evaluation. Their synthesis shows that DL dominates, while traditional ML appears much less often, and that graph-based and token-based representations are the most common inputs. Complementing these accounts, Shimmi et al. \cite{shimmi2025ai} focus on AI-based detection from 2018 to 2023 and propose a taxonomy that consolidates deep techniques, feature representation, and embedding strategies, while underscoring recurring limitations, including dataset quality problems, reproducibility challenges, and weak generalisation to real-world settings.

Domain-specific syntheses extend these streams into particular ecosystems. Senanayake et al. \cite{senanayake2023android} compare static, dynamic, and hybrid methods for Android and report static analysis as the most widely used approach, while also identifying weaknesses in available Android benchmarks and calling for better labelled datasets. Wang et al. \cite{wang2025review} study learning-based detection for smart contracts and conclude that code representation can matter more than classifier choice, comparing sequence, tree, graph, and mixed representations tailored to blockchain semantics. Lin et al. \cite{lin2024vulnerabilities} provide an open-source perspective by surveying how SV and patch detection are linked in OSS projects and by listing the public databases and repositories that underpin detection studies.

A related prediction stream aims to flag artefacts likely to contain vulnerabilities. Bassi and Singh \cite{bassi2023systematic} consolidate model families and predictor sets and enumerate feature types, preprocessing steps, datasets, balancing techniques, and performance measures, thereby systematising the main challenges in developing SV prediction models. Croft et al. \cite{croft2022data} focus specifically on data preparation and develop a taxonomy of problems such as label noise, severe class imbalance, and redundancy from duplicate or versioned code. Kalouptsoglou et al. \cite{kalouptsoglou2023software} situate prediction within a broader map of research goals, dataset construction, model building, and evaluation across 180 studies, clarifying where evidence clusters and how evaluation practices have evolved.

More operational perspectives examine where detection capabilities sit in practice. Fu et al. \cite{fu2025ai} describe how AI-driven security approaches are integrated into DevSecOps workflows, including just-in-time defect prediction at code-commit stages and security checks in CI/CD pipelines. Zhou et al. \cite{zhou2025large} review how LLMs are applied to SV detection alongside repair, describing strategies such as fine-tuning, prompt engineering, and retrieval augmentation, while also noting gaps in high-quality datasets and realistic evaluation. Read together, these reviews show coverage of model families, code representations, data sources, and benchmarking practices, but much less systematic evidence on real-world adoption and workflow integration. That gap motivates the practitioner-centred perspective of our review.

\subsubsection{Assessment}
SV assessment is the evaluation that follows SV detection and precedes SV repair, to determine the characteristics of identified weaknesses and prioritising fixing efforts \cite{dissanayake2022software,le2022survey}. It gathers context about each SV, including severity, exploitability, impact, and type, so remediation decisions can be made on an informed basis \cite{le2022survey}. In practice, this often involves Common Vulnerability Scoring System (CVSS)-based scoring, alongside broader consideration of exploit availability, patch information, and the prevalence of SVs across products \cite{shahzad2019large}. Assessment increasingly relies on data-driven approaches. NLP, ML, and DL are used to automate tasks such as predicting CVSS metrics, exploit likelihood, and SV types, which reduces manual effort and supports faster, more consistent assessment at scale \cite{le2022survey}. 

Le et al. \cite{le2022survey} present a taxonomy of five main themes for data-driven SV assessment and prioritisation: exploitation, impact, severity, type, and miscellaneous tasks. They identify commonly used data sources, for example, NVD and CVE records, vendor advisories, and environmental context, with NVD/CVE being the most prevalently used. They report that ensemble learners such as random forests, XGBoost, and LightGBM often outperform single models. They also highlight practical obstacles, including inconsistencies in SV records, missing relevant external sources, and data quality concerns that can significantly impact model performance.

Elder et al. \cite{elder2024survey} focus on exploitability assessment through three categories: manual CVSS-based, deterministic, and probabilistic assessments. They examine whether disclosed SVs are likely to be exploited and catalogue predictors used across literature, including features from SV reports, social network signals, and exploit availability indicators. They compare scoring and learning approaches, discuss evaluation practices, and warn against feature leakage when inputs contain information unavailable at decision time. However, they note that temporal models predicting exploitation timing performed worse than non-temporal ones, with accuracy decreasing from 89\% to 79\% in one study.

Fu et al. \cite{fu2025ai} examine AI-driven security approaches across DevSecOps stages, identifying 12 security tasks spanning the development life cycle. Within SV assessment, they cover code classification methods that predict SV types, which can be used to prioritise fixes based on the predicted labels. They also discuss SV severity estimation as part of comprehensive security tooling. They describe integration points, including just-in-time defect prediction at commit and continuous integration and continuous delivery (CI/CD) pipeline security checks, which support SV assessment by enabling automated evaluation of code changes for security issues and facilitating prioritised remediation efforts.

For assessment, these reviews explain how evidence flows from technical artefacts to triage decisions and what constitutes an actionable risk signal, while still offering limited coverage of how organisations embed these models at scale or measure their impact on patch scheduling and backlog reduction.

\subsubsection{Repair}
SV repair, also referred to as patching, fixing, or remediation, is the process by which software practitioners develop and release solutions to neutralise identified SVs in their products, with the aim of eliminating the weakness or significantly reducing the associated risk \cite{shahzad2019large,shahzad2012large}. The core phases of patch management are acquiring relevant patches, testing them for compatibility and effectiveness, deploying them to target systems, and confirming successful deployment \cite{dissanayake2022software}. Although repair is the most effective strategy for mitigating SVs, practice is challenged by patch complexity, coordination among stakeholders, and the need to balance timely updates with system availability \cite{dissanayake2022software}. Automation increasingly frames SV repair as code transformation, often as a sequence-to-sequence problem in which a model generates repaired code for a vulnerable snippet \cite{fu2025ai,zhou2025large}. These approaches can integrate with IDEs to provide near real-time suggestions and reduce the time and effort of manual repair \cite{fu2025ai}.

Dissanayake et al. \cite{dissanayake2022software} shift the focus from model performance to patch management as a socio-technical process. They identify fourteen socio-technical challenges across the process, for example, the lack of support for dynamic environment settings where existing approaches are generally "one size fits all", and tools that lack accuracy, security, and scalability. Coordination, communication, and role clarity emerge as persistent barriers; the review also stresses that patch management phases are tightly coupled, where outputs from one phase become inputs to the next. The authors map challenges to reported responses in the literature, classifying approaches, tools, and practices, yet note a substantial evaluation gap; only about 20.8\% of solutions are rigorously assessed in industrial settings.

Lin et al. \cite{lin2024vulnerabilities} connect this process view to concrete artefacts in open-source ecosystems. They outline a SV--patch life cycle that spans six stages: SV generation, discovery, disclosure, exploitation, patch release, and SV demise, and survey how security patches are identified by correlating commits, issues, advisories, and CVE references. Their list of datasets shows language and weakness-type skew, with under 16.7\% of CWE types represented and a concentration on C and C++, and they document the information sources that underpin both detection and patch identification, including NVD, OSV, and vendor bulletins. This ecosystem-level picture clarifies how signals flow from disclosure to patch release and application, and where data coverage limits generalisability.

Fu et al. \cite{fu2025ai} examine AI-driven SV repair as part of the development step in DevOps. They report that AI-driven repair approaches can automatically recommend fixes for vulnerable or buggy programs, reducing the time-consuming and labour-intensive nature of manual code repair. The review covers sequence-to-sequence learning methods, transformer-based pre-trained language models for program repair and SV repair, and specialised techniques such as vulnerability masking to guide repair models. It also identifies end-to-end approaches for both SV detection and repair, demonstrating integrated detection and remediation capabilities in AI-driven security tools.

Zhou et al. \cite{zhou2025large} extend the discussion on automation by reviewing how LLMs are adapted for repair alongside detection. They group adaptation techniques into three categories, fine-tuning, prompt engineering, and retrieval augmentation, and summarise the LLMs and integration patterns used with program analysis. Open issues remain, including a lack of high-quality SV-fix datasets and evaluation designs that rely on synthetic code or real-world fixes without test cases; most studies are evaluated offline on static data and are not integrated into developer workflows.

Read together, these reviews progress from socio-technical challenges, through evidence flows in the SV--patch ecosystem, to the placement of repair within DevSecOps pipelines, and finally to emerging LLM-based repair. Evidence on practitioner evaluation and adoption remains scarce in secondary studies.

\subsubsection{Disclosure}
SV disclosure is the process of making information about a vulnerability publicly available after establishing that it poses a risk \cite{shahzad2019large}. It informs users, stakeholders, and the community about threats and remediation measures, typically through public SV databases and direct communication with affected parties. Disclosure can follow different models, including full disclosure, private disclosure, and coordinated vulnerability disclosure (CVD) \cite{liu2025empirical}.

The timing and manner of disclosure strongly influence security outcomes. Under CVD, issues are reported privately to maintainers and remain confidential until public release, often after a patch is available \cite{liu2025empirical,lin2024vulnerabilities}. In practice, deviations are common, and disclosure may precede patch release or emerge through issue trackers, advisories, or public channels \cite{liu2025empirical}. Disclosure status is also a strong predictor of the time between discovery and exploit publication \cite{elder2024survey}, while SV information may appear on social media before sources such as NVD\footnote{National Vulnerability Database \url{https://nvd.nist.gov/}}, enabling earlier detection but complicating coordinated release \cite{le2022survey}. More broadly, the disclosure ecosystem relies on many information sources, including the CVE\footnote{Common Vulnerabilities and Exposures \url{https://www.cve.org/}} list, NVD, security advisories, vendor notifications, mailing lists, forums, news, blogs, and social media \cite{dissanayake2022software,le2022survey,elder2024survey}. The absence of a single aggregation platform and incomplete reporting complicate monitoring and coordination \cite{dissanayake2022software,croft2022data}.

Table~\ref{tab:SLR-comparison} applies a strict rule for task coverage; a tick indicates explicit, scoped analysis of that task in the review. Under this rule, we did not identify any secondary studies whose primary focus is SV disclosure in SE. Some reviews acknowledge disclosure within broader accounts. For instance, Lin et al. \cite{lin2024vulnerabilities} describe a SV--patch life cycle that includes a disclosure stage when linking advisories and CVEs to patches. However, that review does not examine disclosure processes, stakeholder roles, governance and policy, communication channels, or timing as a dedicated topic. This is why the Disclosure column remains unticked in Table~\ref{tab:SLR-comparison}.

\medskip
\noindent
Overall, the existing reviews offer solid coverage of algorithms, datasets, and evaluation setups for specific SVM tasks, and attention to pipeline placement and ecosystem perspectives is growing. What remains limited at the level of secondary studies is practitioner-centred synthesis, especially analysis that links reported challenges to adoption decisions, workflow integration, and operational outcomes. Our review addresses these gaps by centring studies with industry participants across detection, assessment, repair, and disclosure, by bringing approaches and tools into dialogue with practice-oriented contributions, and by synthesising the challenges that arise when SVM solutions are adopted and used in real-world settings, with explicit attention to context and strength of evidence.

\section{Research methodology}
\label{Sec:Method}
\begin{figure*}[htbp]
  \centering
  \includegraphics[width=0.9\linewidth]{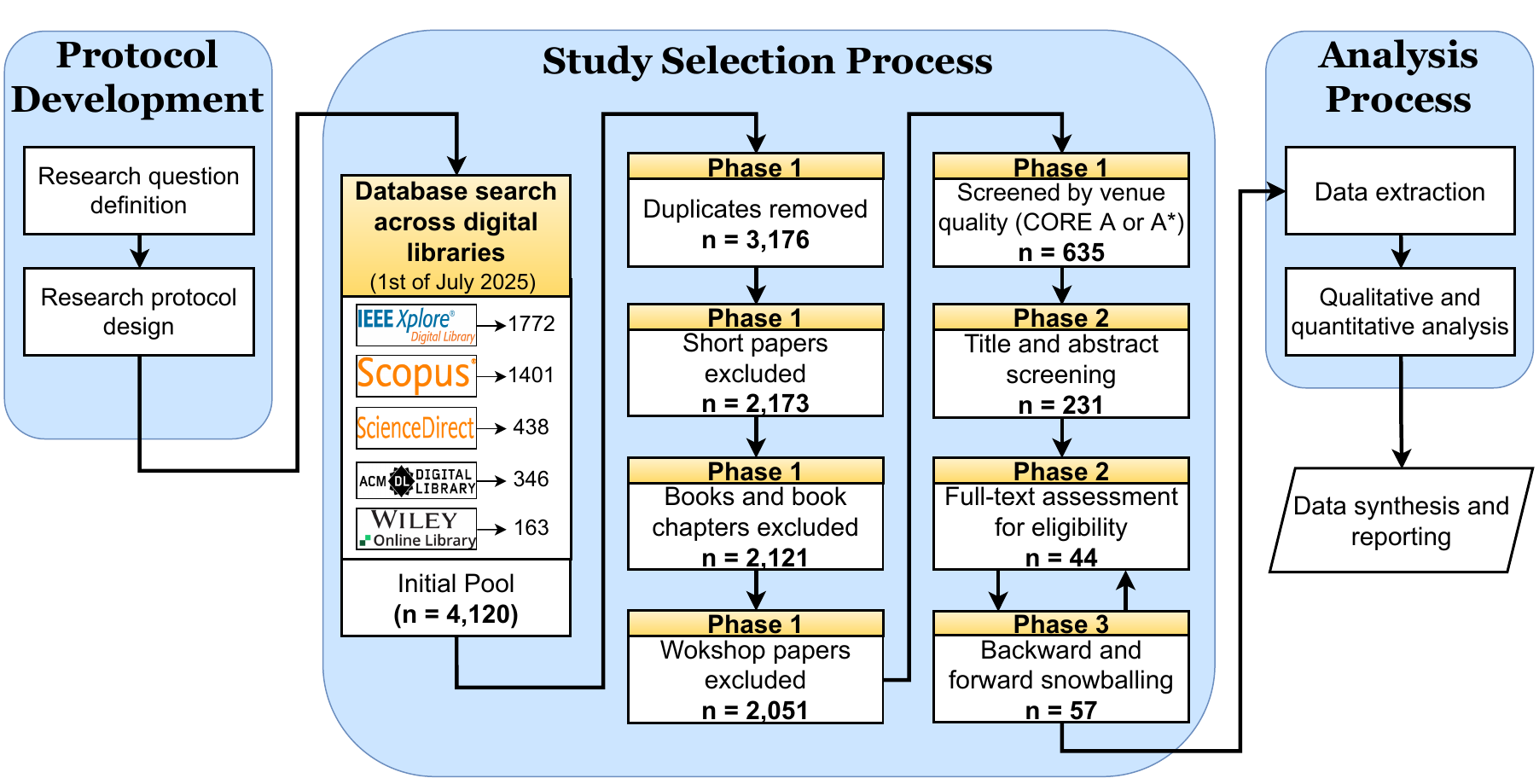}
  \caption{Overview of the research methodology, from research protocol development to data synthesis.}
  \label{fig:Overview}
\end{figure*}

We followed the guidelines of Kitchenham et al. \cite{kitchenham2007guidelines} in designing and executing this SLR. All authors discussed and agreed on the research questions and the protocol through rounds of discussion. The protocol specifies the review objectives, research questions, search strategy, inclusion and exclusion criteria, study selection procedure, and the plan for data extraction and analysis. The overall workflow is shown in Figure \ref{fig:Overview}. The first author performed the database search, then screened studies against the criteria. Backward and forward snowballing complemented the database search to reduce the risk of missing relevant studies. Study selection was performed by the first author and verified by the second author at each stage; disagreements were discussed and resolved, reasons for exclusion were recorded, and the final set of studies was established. For data extraction, the first author conducted a pilot on a subset of studies to validate the extraction form, which the second author reviewed and refined. The first author then completed data extraction and analysis under the supervision of coauthors experienced in conducting SLRs in SE. The team met weekly for approximately six months to review intermediate decisions, resolve uncertainties, and review the evolving analysis. 

\vspace{-5pt}

\subsection{Research questions}
\label{Sec:research-questions}
As this SLR examines SVM from an industry-practitioner perspective, we address two research questions covering the evidence base: the solution space reported with practitioner involvement, challenges practitioners encounter in real-world settings, and how included studies address them. The research questions were defined a priori in the protocol:

\begin{itemize}

\item \textbf{RQ1: What types of solutions for SVM have been proposed and evaluated in studies involving industry practitioners?}

This RQ investigates the solution space for SVM in studies that involve industry practitioners. 
We address this question through two subquestions:

\begin{itemize}
    \item \textit{RQ1.1)} What approaches and tools have been introduced and evaluated with practitioners for SVM?
    \item \textit{RQ1.2)} What practices have been reported or recommended for carrying out SVM in real-world contexts?
\end{itemize}

\item \textbf{RQ2: What challenges do practitioners report when adopting and using SVM solutions in real-world contexts?}

This question explores the barriers practitioners encounter when adopting and using SVM solutions in industry settings. We use thematic analysis to organise practitioner-reported challenges into data-driven themes that synthesise findings across the included studies.

\end{itemize}

Since RQ1 and RQ2 both refer to solutions, we explicitly adopt the categorisation strategy introduced by Shahin et al. \cite{shahin2017continuous} and subsequently used by Dissanayake et al. \cite{dissanayake2022software}. Following their formulation, we use dictionary definitions from the Cambridge and Longman dictionaries, where ``approach'', ``method'', and ``technique'' mean \textit{``a particular way of doing something''}; a ``tool'' is \textit{``something that helps in a particular activity''}; and ``practice'' denotes \textit{``the act of doing something regularly or repeatedly''} \cite{CambridgeDictionary,LongmanDictionary}. In line with their formulation \cite{shahin2017continuous,dissanayake2022software}, we group approaches, methods, techniques, frameworks, algorithms, and implemented tools as technical and formalised means that support SVM activities, referred to as \textbf{\textit{approaches and tools}}. Practices are defined as social practices and shared standards that may be supported by approaches or tools \cite{dittrich2016does,schmidt2014concept}. We classify recommendations, guidelines, best practices, lessons learned, and shared practitioner experiences as \textbf{\textit{practices}} in this review. 

\subsection{Search strategy}
We developed the search strategy to retrieve all relevant literature that meets the inclusion and exclusion criteria in Section~\ref{Sec:inclusion-exclusion}. The design follows established guidelines for systematic reviews \cite{kitchenham2007guidelines} and draws on recommendations for identifying relevant studies in SE \cite{zhang2011identifying}. The strategy comprises two components: selecting the data sources to query and formulating queries adapted to each database’s syntax and fields. The following subsections describe these components.

\subsubsection{Data sources}
We queried (On 1 July 2025) five major digital libraries that are widely used in SE research: IEEE Xplore, Scopus, ScienceDirect, ACM Digital Library, and Wiley Online Library. This selection follows common practice in prior SE reviews and provides broad coverage across publishers and indexing services \cite{shahin2014systematic,shahin2017continuous}. We did not include Google Scholar, since prior assessments report low precision, many irrelevant results, and substantial overlap with ACM Digital Library and IEEE Xplore for SE literature \cite{chen2010towards}. Automatic searches were run in each source using the search terms defined in Section~\ref{Sec:search-terms}, and the retrieved records were pooled, then deduplicated. We complemented the automatic search with backward and forward snowballing \cite{wohlin2014guidelines}, as explained in Section~\ref{Sec:study-selection}.

\begin{tcolorbox}[colback=gray!10, colframe=black, title=Search string]
(\texttt{``developer*'' OR ``programmer*'' OR ``practitioner*'' OR ``professional*'' OR ``specialist*'' OR ``expert*'' OR ``engineer*'' OR ``analyst*'' OR ``tester*'' OR ``responder*'' OR ``architect*'' OR ``auditor*'' OR ``officer*''}) \textbf{AND} (\texttt{``vulnerabilit*'' OR ``defect*'' OR ``bug*'' OR ``flaw*'' OR ``weakness*'' OR ``fault*'' OR ``threat*''}) \textbf{AND} (\texttt{``user study'' OR ``interview*'' OR ``survey*'' OR ``questionnaire*'' OR ``empirical stud*''}) \textbf{AND} (\texttt{``software*'' OR ``cod*'' OR ``tool*'' OR ``practice*'' OR ``secur* analys*'' OR ``secur* test*''}) \textbf{AND} \texttt{``secur*''}
\end{tcolorbox}

\subsubsection{Search terms}
\label{Sec:search-terms}
We constructed the search string to retrieve empirical studies involving practitioners on SVM. 
According to the guidelines by Kitchenham et al. \cite{kitchenham2007guidelines}, we used the following strategies to form the search string. 
First, we identified the main concept groups based on our research questions. Next, we compiled a list of common keywords for each group, using a small set of relevant papers that the team was familiar with. We also included close synonyms for these keywords and conducted small pilot searches to ensure that the string retrieved the known papers. 

Guided by the research questions, we used four concept groups. The first group enumerates practitioner roles to focus on industry participants (e.g., developer, analyst). The second group covers the terms used for vulnerabilities in software contexts, including synonyms and colloquial terms such as bug and fault. The third signals empirical methods with practitioners (e.g., interview, survey). We also included the search term \texttt{``empirical stud*''} because we noticed many studies where the authors do not specify the exact empirical method (e.g., interviews) used in the study within the abstract section. Instead, they only mention the term ``empirical''. We took this approach to ensure we do not overlook any relevant studies in this area. The fourth anchors the scope to software and development work (e.g., software, code), so results focus on software artifacts and activities. The final \texttt{AND ``secur*''} then narrows the set to security-relevant studies.

We applied the string to titles, abstracts, and author keywords where supported by each database. No date limits were applied. Wildcards were used in all databases except ScienceDirect to cover a large number of alternative search terms. In the ScienceDirect advanced search, only eight operators (e.g., AND and OR) are allowed per query. Therefore, we issued combinations of the search terms to keep the queries consistent across databases and then merged the results for screening. Beyond these operator limits, we made small syntax adjustments per database, such as quoting phrases or changing field scoping, while keeping the same Boolean structure. 

\subsection{Inclusion and exclusion criteria}
\label{Sec:inclusion-exclusion}

\begin{table}[htbp]
\centering
\caption{Inclusion and exclusion criteria}
\label{tab:inclusion-exclusion}
\begin{tabular}{p{0.10\linewidth} p{0.85\linewidth}}
\toprule
\textbf{ID} & \textbf{Inclusion criteria} \\
\midrule
IC-1 & The study empirically evaluates, applies, or investigates an approach, tool, or practice targeting at least one phase of the SVM life cycle. \\
IC-2 & The study collects primary data directly from industry practitioners (e.g., developers, security engineers) through methods such as interviews, questionnaires, or field observations. \\
IC-3 & The paper is published in a peer-reviewed venue ranked A* or A by the CORE ranking. \\
IC-4 & The paper is written in English and is accessible in full text. \\
\midrule
\textbf{ID} & \textbf{Exclusion criteria} \\
\midrule
EC-1 & The paper is a short paper with fewer than six pages. \\
EC-2 & The paper is published as a workshop article, dissertation, book, or other non-peer-reviewed work such as technical reports, keynotes, tutorials, or panel discussions. \\
EC-3 & The paper is a review or survey paper (e.g., literature reviews). \\
EC-4 & The paper includes student participants or relies exclusively on secondary data sources (e.g., repository mining, forum content). \\
\bottomrule
\end{tabular}
\end{table}

Table~\ref{tab:inclusion-exclusion} lists the criteria used to filter all studies retrieved from the digital libraries. To be included, a study needed to empirically evaluate, apply, or investigate an approach, tool, or practice that targets at least one phase of the SVM life cycle (IC-1). It also needed to collect primary data directly from industry practitioners through methods such as interviews, questionnaires, or field observations (IC-2). Venue quality was operationalised using the CORE ranking\footnote{CORE Portal \url{https://www.core.edu.au/conference-portal}}, and we kept articles published in A* or A venues \cite{GoogleDrive_Descriptions} only (IC-3). We adopted this venue assessment criterion to focus on high-quality articles. The CORE ranking combines expert judgement with empirical indicators drawn from venue activity and impact, for example, citation patterns of published papers, involvement of established researchers, acceptance rates, and the track record of program chairs. The list is reviewed and updated regularly \cite{COREConferencePortal}. Although developed in Australia, the methodology relies on widely used indicators and is applied internationally, and it has been used to screen venues in many SE systematic reviews \cite{fu2025ai,croft2022data,sarker2024multi,mousavi2025detecting}.
Only peer-reviewed studies in English with full text available were considered (IC-4).

We excluded short papers under six pages (EC-1), workshop papers, dissertations, books, and other non-peer-reviewed material such as technical reports, keynotes, tutorials, or panel discussions (EC-2). We also excluded review or survey papers (EC-3), as well as studies that involved student participants or relied exclusively on secondary data, such as repository mining or forum content (EC-4).

\subsection{Study selection}
\label{Sec:study-selection}
Figure~\ref{fig:Overview} summarises the study selection workflow and shows the number of studies at each step. The predefined inclusion and exclusion criteria in Section~\ref{Sec:inclusion-exclusion} guided every decision. We kept a screening log with titles, decisions, and reasons for exclusion so that the process is transparent and reproducible. The first author carried out the screening, the second author verified all decisions, and any disagreements were resolved through discussion. The study selection process comprises the database search followed by three screening phases.

\textbf{Database search:} On 1 July 2025, we ran the search string across the five digital libraries in Section~\ref{Sec:search-terms} and formed a pool of 4{,}120 studies. All available metadata and identifiers were exported to support screening and deduplication.

\textbf{Phase 1:} We began by removing duplicates across sources, reducing the pool to 3{,}176 studies. We then filtered studies by length and publication format using the exclusion criteria. Short papers under six pages were removed, leaving 2{,}173 studies. Books and book chapters were removed, leaving 2{,}121 studies, followed by workshop papers, leaving 2{,}051. Finally, we screened venues using the CORE ranking and retained only A* and A venues, yielding 635 studies.

\textbf{Phase 2:} We screened titles and abstracts against the inclusion criteria (IC-1 and IC-2). If a decision could not be made on the basis of title and abstract, the study was carried forward so that potentially relevant work was not lost. After this screen, 231 studies remained. We then read the full text of these studies and applied all criteria. Reasons for exclusion were recorded (e.g., no clear connection to the SVM life cycle). This phase concluded with 44 studies.

\textbf{Phase 3:} We conducted backward and forward snowballing on the Phase 2 set, following the standard procedure \cite{wohlin2014guidelines}, to reduce the risk of missing relevant studies. For forward snowballing, we used Google Scholar to identify papers citing the Phase 2 studies. For backward snowballing, we examined their reference lists. All newly found studies were processed using the same sequence, including deduplication, length and format checks, venue screening, title and abstract screening, and full-text assessment. After applying the criteria, the final set comprised 57 studies in total.

\subsection{Data extraction}
\label{Sec:data-extraction}
We extracted data from the included studies using a structured form in a spreadsheet. Table~\ref{tab:data-extraction} shows the extraction form with the data items and their links to the research questions. The first author conducted a pilot extraction on six randomly selected papers, approximately 10\% of the total, to check that the extracted fields would support answering the RQs. The pilot indicated that the form captured the main study characteristics but that some items, such as Area of focus (D7) and Data collection methods (D9), required clearer definitions and guidance for consistent coding. The second author reviewed the pilot output and suggested refinements, after which we fixed the schema to maintain consistency. In line with established SLR guidelines, data extraction involved more than one author \cite{kitchenham2004procedures}. After the pilot, the first author continued the extraction in batches with regular checkpoints. To reduce bias and drift during extraction, the team reviewed the sheet in regular meetings and resolved ambiguities through discussion. The completed extraction sheet forms the basis for the analyses reported in Section~\ref{Sec:data-analysis}.

The form contains three groups of items, defined following recommended data extraction practices for SLRs in software engineering \cite{garousi2017experience, kitchenham2004procedures}. \textit{Bibliographic} items (D1--D5) record title, authors, venue, year, and publication type. \textit{Contextual attributes} (D6--D9) capture application domain, SVM phase or phases addressed, participant roles, and the data collection methods used with practitioners. These items do not answer the RQs directly, but they complement the RQ-focused items by allowing us to describe the corpus and analyse trends (e.g., which SVM phases or domains are most studied). \textit{RQ-focused} items (D10--D13) capture the substantive evidence used to answer the research questions. We first classified each study's contribution as \textit{approaches and tools} or \textit{practices} (D10) to structure the evidence for RQ1. For RQ1.1, we recorded approaches and tools introduced for SVM and evaluated with practitioners (D11). For RQ1.2, we recorded practices reported or recommended for conducting SVM in real settings (D12). For RQ2, we captured practitioner-reported challenges to adopting or using SVM solutions (D13). 

\begin{table*}[htbp]
\centering
\caption{Data extraction form}
\label{tab:data-extraction}
\small
\begin{tabular}{p{0.04\linewidth} p{0.19\linewidth} p{0.45\linewidth} p{0.16\linewidth}}
\toprule
\textbf{ID} & \textbf{Data item} & \textbf{Description} & \textbf{Analytic category} \\
\midrule
D1  & Title                      & Title of the paper. & Bibliographic \\
D2  & Author(s)                  & Authors of the paper. & Bibliographic \\
D3  & Venue                      & Publication venue. & Bibliographic \\
D4  & Year                       & Year of publication. & Bibliographic \\
D5  & Publication type           & Type of publication (conference, journal). & Bibliographic \\
\midrule
D6  & Application domain         & Target domain (e.g., smart contracts). & Contextual attributes \\
D7  & Area of focus              & SVM phase or phases addressed. & Contextual attributes \\
D8  & Participant roles          & Practitioner roles involved (e.g., developer, security engineer). & Contextual attributes \\
D9 & Data collection methods    & Methods with practitioners (e.g., interview, survey). & Contextual attributes \\
\midrule
D10  & Solution type              & Classification of the contribution as \textit{approaches and tools} or \textit{practices}. & RQ1 (classification) \\
D11 & Approaches and tools       & \textit{Approaches and tools} introduced for SVM and evaluated with practitioners. & RQ1.1 \\
D12 & Practices                  & \textit{Practices} reported or recommended for conducting SVM in real-world contexts. & RQ1.2 \\
D13 & Challenges                 & Practitioner-reported challenges when adopting or using SVM solutions. & RQ2 \\
\bottomrule
\end{tabular}
\end{table*}

\subsection{Data analysis and synthesis}
\label{Sec:data-analysis}
Bibliographic items (D1--D5), contextual attributes (D6--D9), and the solution type classification (D10) were analysed using descriptive statistics, and we reported frequency distributions and simple cross-tabulations to show patterns in the corpus. For RQ1, the solution-focused items (D11 and D12) were analysed using descriptive and qualitative mapping. We grouped studies by contribution type (e.g., tool, framework, method), SVM phase focus (i.e., Detection, Assessment, Repair, Disclosure), and we synthesised the findings narratively. This analysis underpins the overviews of approaches and tools (RQ1.1) and practices (RQ1.2) reported in Section~\ref{Sec:Result}.

For RQ2, the challenge-focused item (D13) was analysed using Braun and Clarke’s method for thematic analysis \cite{braun2006using} to organise and interpret the qualitative evidence. We used this method because it is widely adopted in SLRs in SE \cite{shahin2014systematic,dissanayake2022software,sarker2024multi,shahin2017continuous,croft2022data,mousavi2025detecting}. For the qualitative synthesis of RQ2, we used an inductive, data-driven strategy. The unit of analysis was text excerpts taken from the included studies and recorded for the challenge-related item (D13) in Section~\ref{Sec:data-extraction} (e.g., practitioner-reported challenge statements, accompanied by brief explanations of why practitioners find them challenging or what consequences they report). The RQ2 challenge dataset was analysed using a codebook whose codes were organised into subthemes (ST) and themes (T). We first familiarised ourselves with all excerpts and revisited source papers when clarification was needed. We then ran a pilot in which we randomly selected 6 papers, 10\% of the total, and the first and second authors independently coded the challenge-related excerpts and produced initial codebooks. The two codebooks were compared and reconciled through discussion, and overlapping codes were merged into a consolidated version. Using the consolidated codebook, the first author coded the remaining excerpts while documenting any new codes that needed to be developed.

We grouped related codes into subthemes and organised them into broader themes. Each subtheme and theme was given a name and definition. We maintained explicit links from each excerpt to its code, subtheme, and theme.

To improve reliability, the first and second authors independently coded the 6-paper pilot sample, compared results, and reconciled differences before full coding. During full coding, the team met weekly to discuss emerging codes, subthemes, and themes and to ensure consistent application of definitions. 
The final themes, subthemes, and mappings provide the basis for the results reported for RQ2 in Section~\ref{Sec:Result}.

\subsection{Overview of the selected studies}
\label{Sec:overview-studies}

\begin{figure}[htbp]
  \centering
  \begin{minipage}[t]{0.49\textwidth}
    \centering
    \includegraphics[width=\linewidth]{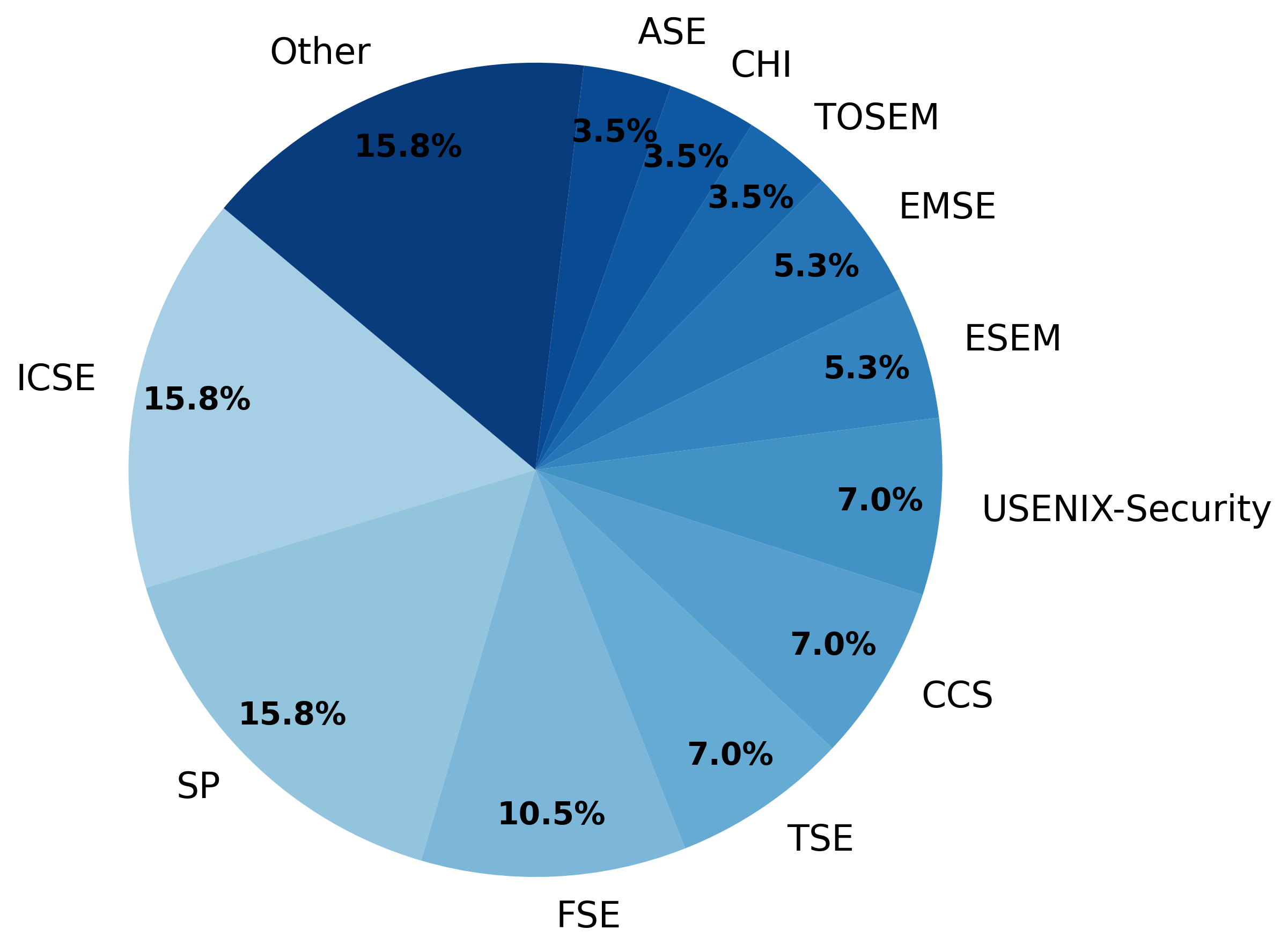}
    \caption{Distribution of papers across individual SE and security venues (D3).}
    \label{fig:venue-distributions-a}
  \end{minipage}\hfill
  \begin{minipage}[t]{0.49\textwidth}
    \centering
    \includegraphics[width=\linewidth]{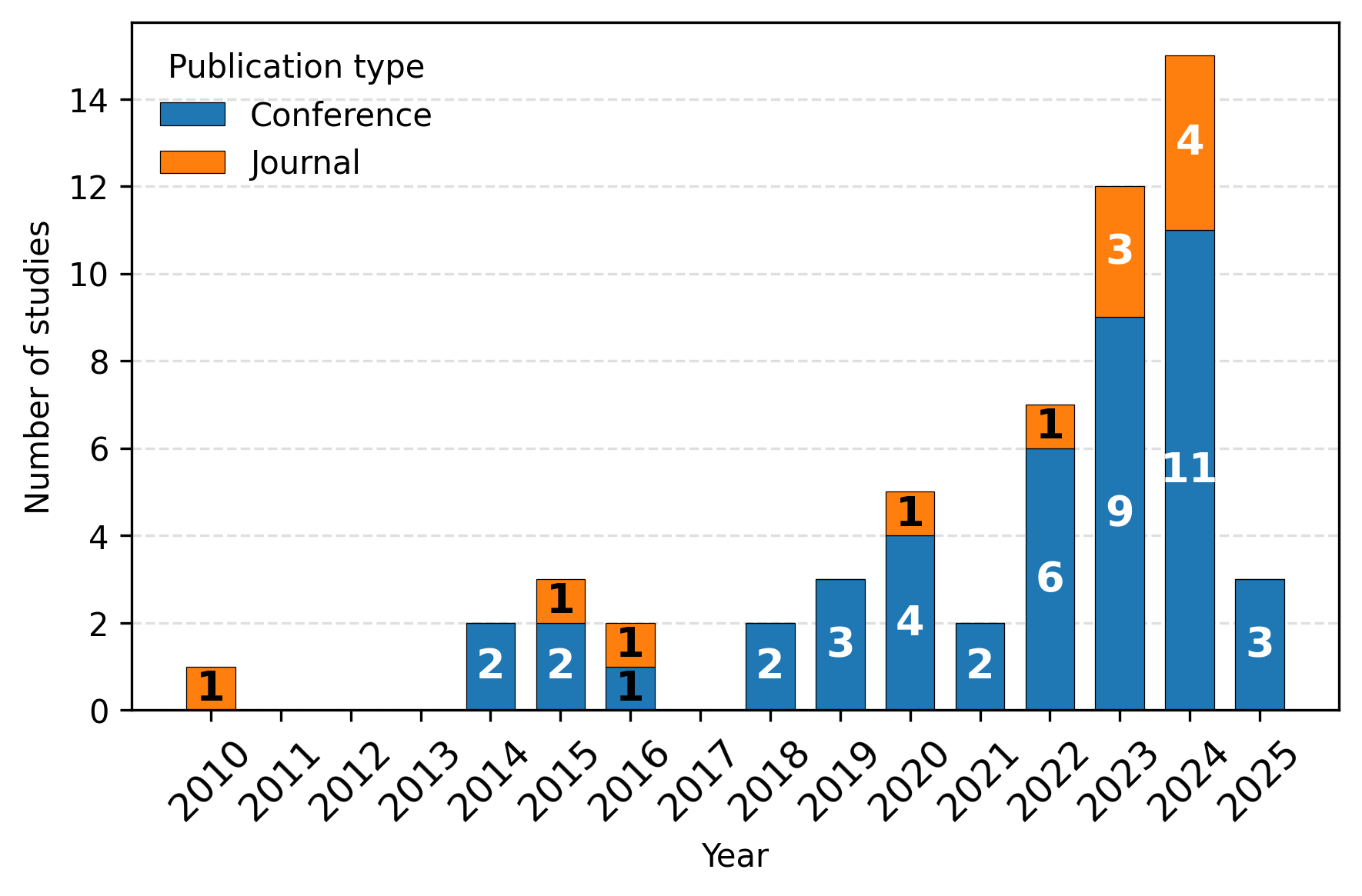}
    \caption{Distribution of studies over years and types of venues (D4 and D5).}
    \label{fig:year-venue-type}
  \end{minipage}
\end{figure}

\textbf{Venue (D3).}
Figure~\ref{fig:venue-distributions-a} shows that the 57 studies are concentrated in a small set of flagship venues. ICSE and the IEEE Symposium on Security and Privacy (SP) each contribute nine papers (16\% per venue), and FSE adds six papers (11\%). A smaller set of venues contributes four papers each (7\% per venue), while the remaining studies are spread across a longer tail of SE and security outlets. Overall, 37 studies (65\%) are published in SE venues and 20 studies (35\%) in security venues, indicating that most practitioner-oriented SVM work is communicated through mainstream SE outlets while security venues still contribute a substantial share. In terms of venue quality, 42 studies (74\%) are published in A* venues and 15 studies (26\%) in A venues according to the CORE ranking system \cite{COREConferencePortal}.

\textbf{Year and Publication type (D4 and D5).}
Figure~\ref{fig:year-venue-type} shows the temporal distribution of the 57 primary studies and distinguishes between conference and journal publications. The earliest study in our corpus was published in 2010, with only sporadic publications until 2018. From 2019 onward, the volume of work increases steadily, with a sharp rise from 2022 to 2024, when 34 studies (60\% of the corpus) were published. A closer inspection of the papers sorted by year suggests that this surge coincides with advances in generative AI and heightened attention to software security risks. The peak occurs in 2024 with 15 studies. Overall, most papers appear in conference proceedings (45 studies, 79\%), while journals account for 12 studies (21\%). This pattern suggests that practitioner-centred work on SVM is an active and growing topic primarily disseminated through top-tier conferences, with journal publications increasing more recently.

The full list of primary studies, including their bibliographic items (D1--D5) and contextual attributes (D6--D9), along with additional breakdowns, is provided in the online appendix \cite{svm-slr-online-appendix}.

\section{Results}
\label{Sec:Result}

\subsection{RQ1}
RQ1 asks what kinds of SVM solutions have been proposed and evaluated in studies involving industry practitioners. In this SLR, we use “solution” as an umbrella term covering technical \textit{\textbf{approaches and tools}} and \textit{\textbf{practice}}-oriented contributions. We operationalise this distinction via Solution type (D10), reported in Section~\ref{Sec:Solution_Type}. Accordingly, we answer RQ1 through two subquestions: RQ1.1 analyses 15 studies that introduce and evaluate SVM approaches or tools with practitioners, and RQ1.2 analyses 42 studies that report or recommend SVM practices in real-world contexts.

\subsubsection{Solution type (D10)}
\label{Sec:Solution_Type}
The 57 primary studies (all involving industry practitioners) report two broad solution types. In line with our definitions in Section~\ref{Sec:research-questions}, we distinguish between \textbf{\textit{approaches and tools}} as technical and formalised means that support SVM activities, and \textbf{\textit{practices}} as social practices and shared standards. Most studies contribute practices, with 42 papers (74\%) reporting recommendations, guidelines, best practices, lessons learned, or shared practitioner experiences. The remaining 15 studies (26\%) focus on approaches and tools, including implemented tools, frameworks, methods, algorithms, and techniques evaluated with practitioners. This distribution suggests that practitioner-centred SVM research to date has placed greater emphasis on capturing and systematising practice-oriented knowledge than on the design and evaluation of concrete technical solutions.

Table~\ref{tab:solution-at-sub} shows the corresponding subcategories for approaches and tools. Implemented tools account for nearly half of the 15 technical contributions, with 7 studies reporting concrete tool support that was evaluated with practitioners. The remaining papers are split between frameworks and methods, with 3 studies in each group, and a small number of standalone algorithms and techniques (1 study each). This pattern suggests that when technical solutions are reported with practitioner involvement, they are most often delivered as implemented tools, while more abstract frameworks and individual algorithms or techniques are less frequently represented.
\begin{table}[htbp]
\centering
\caption{Approach and tool subcategories (D10).}
\label{tab:solution-at-sub}
\small
\begin{tabular}{p{0.42\linewidth} c p{0.35\linewidth}}
\toprule
\textbf{Approaches and tools subcategory} & \textbf{\#Papers} & \textbf{References} \\
\midrule
Implemented tool & 7 &
\cite{Luo2021,Pan2022,Wickert2024,Falsina2015,Holm2015,Steenhoek2025,Fu2024a} \\
\midrule
Framework & 3 &
\cite{Zhou2023,Fu2024,LariosVargas2023} \\
\midrule
Method & 3 &
\cite{Santos2019,Manadhata2010,Venson2023} \\
\midrule
Algorithm & 1 &
\cite{Pan2023} \\
\midrule
Technique & 1 &
\cite{Liu2024} \\
\bottomrule
\end{tabular}
\end{table}

Table~\ref{tab:solution-practice-sub} breaks down the practice-oriented solutions into subcategories. Guidelines are the most common form, with 14 of 42 practice papers. These provide concrete, prescriptive suggestions for how SVM activities should be carried out. Lessons learned, reported in 10 studies, emphasise insights from past projects, highlighting pitfalls, trade offs, and factors that helped or hindered SVM work. Recommendations appear in 7 studies and typically offer targeted, forward-looking advice for stakeholder groups, such as tool builders, managers, or platform providers. Best practices, reported in 6 studies, describe mature ways of working presented as desirable standards organisations should institutionalise. Finally, shared practitioner experiences in 5 studies present rich narrative accounts of SVM work in context, where the primary contribution is to document real-world practices and constraints rather than codify them into formal guidance.
\begin{table}[htbp]
\centering
\caption{Practice-oriented solution subcategories (D10).}
\label{tab:solution-practice-sub}
\small
\begin{tabular}{p{0.42\linewidth} c p{0.35\linewidth}}
\toprule
\textbf{Practice subcategory} & \textbf{\#Papers} & \textbf{References} \\
\midrule
Guideline & 14 &
\cite{Votipka2020,Votipka2018,Mink2023,Fulton2023,Ryan2023,Lopez2022,Nguyen2020,Christakis2016,Thomas2018,Braz2022a,Chaliasos2023,Mandal2025,Liu2024a,Kloeg2024} \\
\midrule
Lesson learned & 10 &
\cite{Weir2020,Hoeltervennhoff2023,Ami2024,Zhang2023,VanDerLinden2020,Venson2019,Andreina2024,Sharma2024,Braz2022b,Assal2019} \\
\midrule
Recommendation & 7 &
\cite{Happe2023,Dissanayake2022,Wunder2024,Braz2022,Hafiz2016,Wan2024,Jancar2022} \\
\midrule
Best practice & 6 &
\cite{Hantke2024,Witschey2015,Wan2021,TenNapel2025,Klostermeyer2024,Xiao2014} \\
\midrule
Shared experience & 5 &
\cite{DiazFerreyra2024,Fang2014,Rebatchi2024,Pashchenko2020,Wu2023} \\
\bottomrule
\end{tabular}
\end{table}

\subsubsection{RQ1.1}
RQ1.1 focuses on \textit{\textbf{approaches and tools}} that have been introduced for SVM and evaluated with industry practitioners. To address it, we analysed the 15 primary studies that proposed an SVM-related approach or tool and evaluated it with practitioners. The full study-level breakdown of the 15 approaches and tools, including their purpose, contribution type, and explicit SVM phase focus, is provided in the online appendix \cite{svm-slr-online-appendix}.

\begin{figure}[htbp]
  \centering
  \begin{minipage}[t]{0.48\textwidth}
    \centering
    \includegraphics[width=\linewidth]{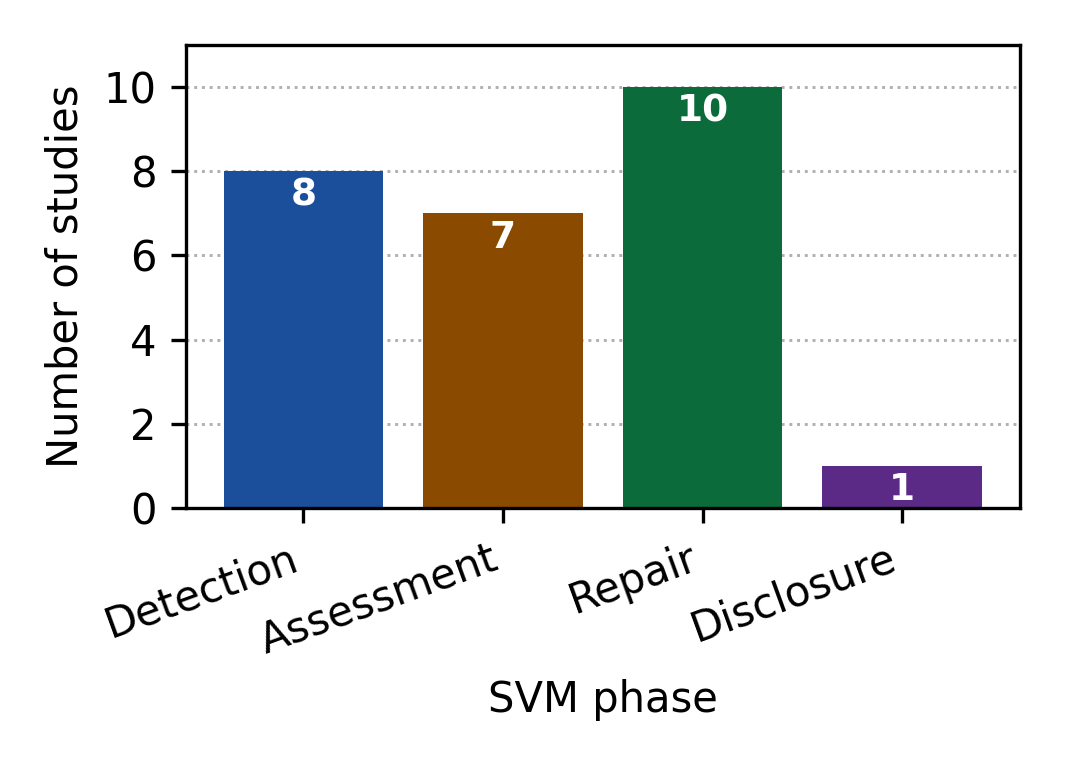}
    \caption{Count of studies by SVM phase focus for approaches and tools (RQ1.1). \textbf{Note:} Studies may contribute to multiple phases.}
    \label{fig:rq11-phase-counts}
  \end{minipage}\hfill
  \begin{minipage}[t]{0.48\textwidth}
    \centering
    \includegraphics[width=\linewidth]{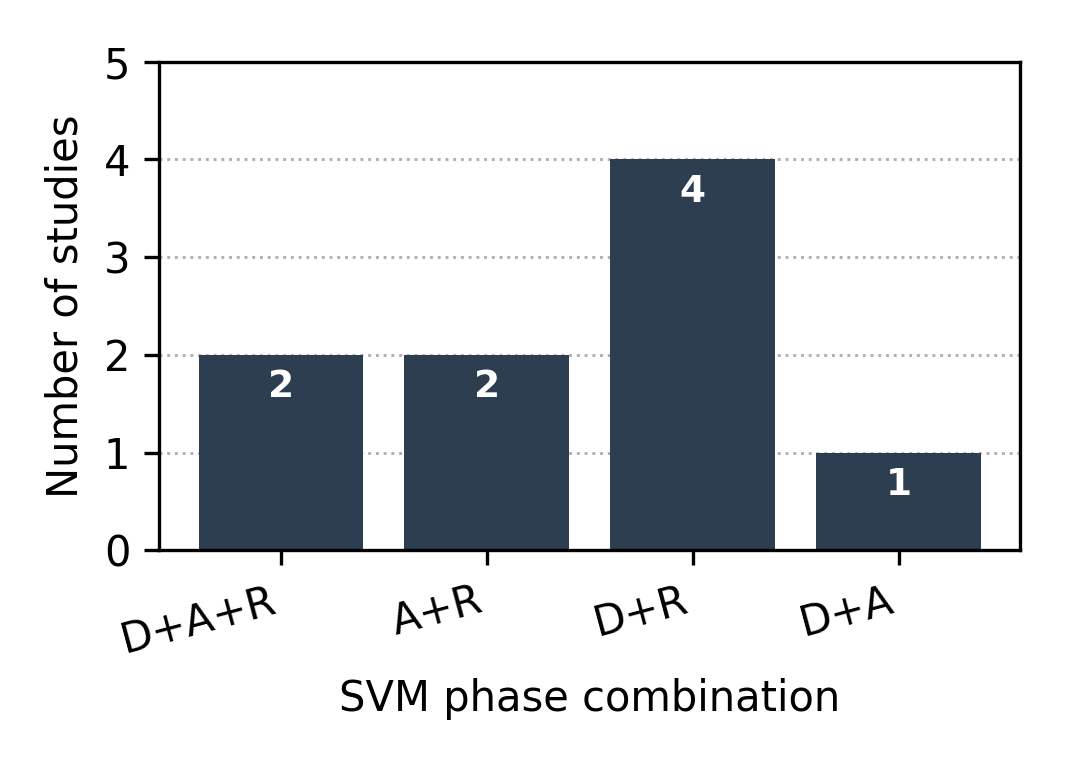}
    \caption{Count of studies by multi-phase combinations for approaches and tools (RQ1.1).}
    \label{fig:rq11-combo-counts}
  \end{minipage}
\end{figure}

\paragraph{\textbf{SVM life-cycle coverage of approaches and tools}}
Figures~\ref{fig:rq11-phase-counts} and~\ref{fig:rq11-combo-counts} summarise the distribution of these approaches and tools across the SVM life cycle and the most common multi-phase combinations. Repair is the most frequently supported phase (10/15), followed by Detection (8/15) and Assessment (7/15), while Disclosure is explicitly targeted by only one study (1/15). Detection+Repair is the most common multi-phase combination (4 studies), followed by Assessment+Repair (2 studies) and Detection+Assessment+Repair (2 studies). Overall, the practitioner-evaluated approaches and tools are concentrated on discovering, prioritising, and fixing SVs, and many span more than one phase of the SVM life cycle. By contrast, Disclosure remains largely unsupported. This pattern suggests that the current solution space places much greater emphasis on detection, assessment, and repair than on managing disclosure activities.

\paragraph{\textbf{Approaches and tools for SV detection}}
Detection support is mainly delivered through developer-facing tooling that surfaces SVs during development and review. This includes IDE extensions and static-analysis enhancements that flag security issues within developers' workflows [S3, S14, S40, S43]. Beyond such tools, detection is also addressed through a grounded-theory-based method uncovering recurring architectural vulnerability patterns in plug-and-play architectures [S2], a behavioural framework for introducing secure coding tools and processes in practice [S28], and a fuzzing-based technique that exercises kernel error behaviours to expose high-risk bugs [S38]. In addition, one method contributes decision support by modelling how required security levels influence development effort [S47].

\paragraph{\textbf{Approaches and tools for SV assessment and prioritisation}}
Assessment and prioritisation are addressed through approaches that classify SVs, estimate severity or exploitability, and support triage decisions. Examples include an ML-based classifier that predicts fine-grained CWE types for SV-fixing commits to support fine-grained triage [S19], a Bayesian-network-based tool that estimates exploit likelihood for critical SVs [S30], and a fuzzing-based technique that helps prioritise kernel bugs with high exploitation potential [S38]. Assessment also appears in approaches that help practitioners understand and prioritise unreported or hard-to-observe security work, such as explainable learning-based support for identifying silent SV fixes and rating their severity [S10], and IDE tooling that detects, classifies, and estimates severity while also suggesting repairs [S43]. Attack-surface measurement provides another assessment-oriented input by quantifying exposure across entry points and channels to inform hardening priorities [S27]. Overall, the assessment-oriented contributions are tightly coupled to triage and prioritisation, often feeding into repair decisions.

\paragraph{\textbf{Approaches and tools for SV repair and patching}}
Repair is the best-supported phase, spanning direct repair generation and repair guidance. Several approaches provide automated or AI-assisted repair capabilities, including a vision-transformer-inspired framework for C/C++ SVs [S21] and an IDE extension that detects issues and recommends fixes during real-world development [S40], alongside editor-integrated tooling that suggests repairs as part of detection and assessment support [S43]. Other work provides repair-enabling mechanisms and safeguards, such as enforcing secure dynamic code loading on Android at runtime [S25] and static-analysis enhancements that support detecting and fixing multi-step cryptographic API misuses [S14]. Repair is supported indirectly through approaches that guide remediation planning and prioritisation, including architectural analysis of vulnerability patterns [S2], attack-surface measurement to prioritise hardening [S27], and support for following up on silent SV fixes [S10]. Taken together, these contributions emphasise repair execution and decision support, often integrated with detection and assessment.

\paragraph{\textbf{Approaches and tools for SV disclosure}}
Only one practitioner-evaluated approach explicitly targets Disclosure. This tool uses a DL-based approach to scan issue reports before disclosure and detect reports that leak SV information, supporting safer reporting and coordinated disclosure practices [S6]. The scarcity of disclosure-focused approaches suggests that disclosure workflows remain largely unsupported by tools that have been evaluated with practitioners.

\begin{center}
\fbox{%
\begin{minipage}{0.96\linewidth}
\textbf{Summary of RQ1.1 (approaches and tools).}

\begin{itemize}
    \item 15 primary studies introduce and evaluate SVM approaches or tools with industry practitioners.
    
    \item The solution space is dominated by developer-facing tools with fewer techniques or standalone algorithms.
    
    \item Most approaches and tools support detection, assessment, and repair, and many span more than one SVM phase. Only one study explicitly targets disclosure, so this phase remains largely unsupported by practitioner-evaluated solutions.
\end{itemize}
\end{minipage}%
}
\end{center}

\subsubsection{RQ1.2}

RQ1.2 asks what \textbf{\textit{practices}} have been reported or recommended for carrying out SVM in real-world contexts. To address this question, we analysed the 42 primary studies that report or recommend practice-oriented solutions (i.e., guidelines, lessons learned, recommendations, best practices, and shared experience reports) for SVM in industry. The full study-level breakdown of the 42 practice-oriented studies, including their SVM practice focus, practice type, and explicit SVM phase focus, is provided in the online appendix \cite{svm-slr-online-appendix}.

\begin{figure}[htbp]
  \centering
  \begin{minipage}[t]{0.45\textwidth}
    \centering
    \includegraphics[width=\linewidth]{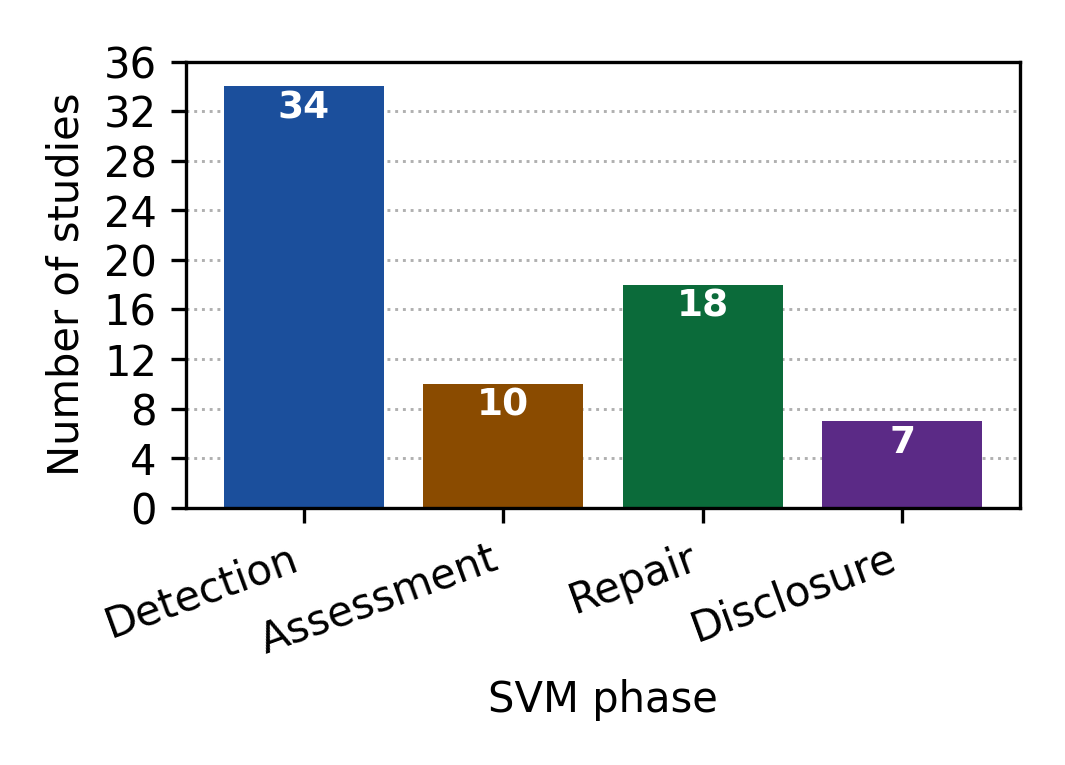}
    \caption{Count of studies by SVM phase focus for practices (RQ1.2). \textbf{Note:} Studies may contribute to multiple phases.}
    \label{fig:rq12-phase-counts}
  \end{minipage}\hfill
  \begin{minipage}[t]{0.45\textwidth}
    \centering
    \includegraphics[width=\linewidth]{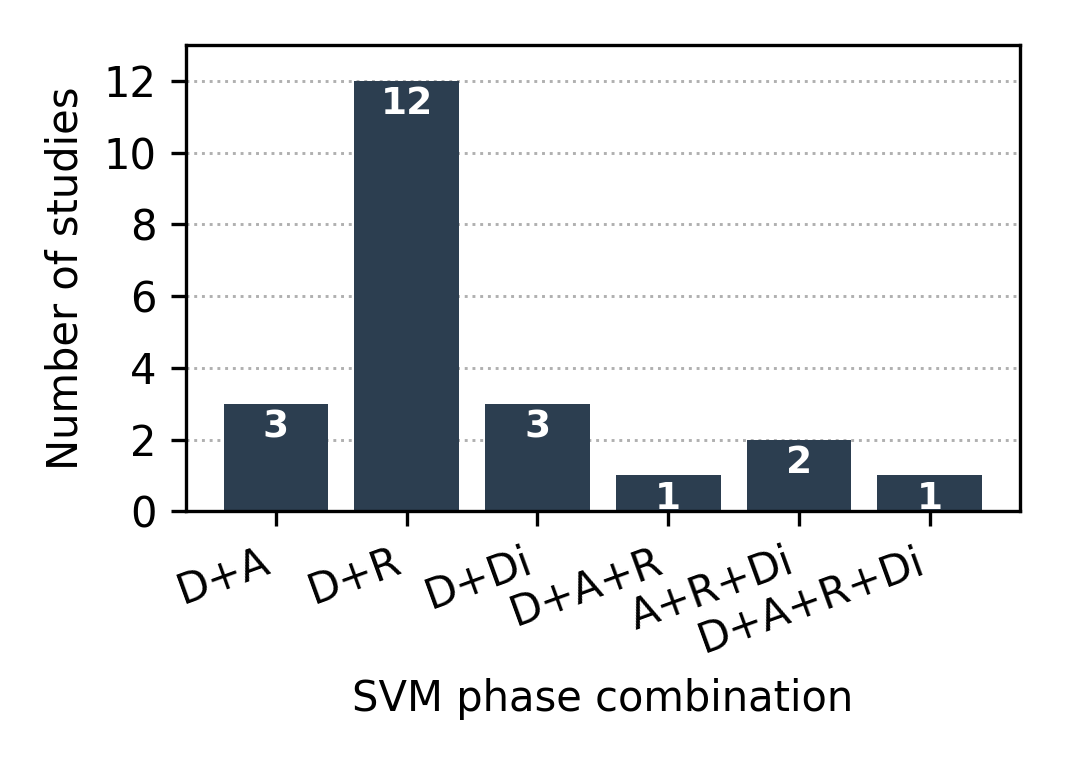}
    \caption{Count of studies by multi-phase combinations for practices (RQ1.2).}
    \label{fig:rq12-combo-counts}
  \end{minipage}
\end{figure}

\paragraph{\textbf{SVM life-cycle coverage of practices}}
Figures~\ref{fig:rq12-phase-counts} and~\ref{fig:rq12-combo-counts} summarise the distribution of the practice-oriented studies across the SVM life cycle and the most common multi-phase combinations. Detection is the most frequently addressed phase (34/42), followed by Repair (18/42), Assessment (10/42), and Disclosure (7/42). Detection+Repair is the most common multi-phase combination (12 studies), followed by Detection+Assessment and Detection+Disclosure (3 studies each), and Assessment+Repair+Disclosure (2 studies). Overall, the practice-oriented studies span all four phases of the SVM life cycle, but they are concentrated on development-time work, especially on finding and fixing SVs. Assessment appears less often as a distinct practice area and is more commonly embedded within broader activities, while Disclosure receives the least explicit attention. Taken together, the evidence suggests that reported practices focus more on how teams handle SVs during everyday development than on providing end-to-end guidance across the full SVM life cycle.

\paragraph{\textbf{Practices for SV detection}}
Detection-oriented practices centre on how teams incorporate security into development work. Several studies examine secure coding behaviours and security culture [S4, S18, S22, S55], showing how requirements, communication, and organisational expectations influence whether SVs are noticed during development. Others focus on developers working with security tooling to surface SVs, including static and program analysis tools [S12, S15, S29, S31] and ML- or AI-based SVM tools integrated into workflows [S12, S49]. Code review emerges as a second focus of detection work, where reviewers rely on cues and review strategies to identify SVs in change sets [S23, S34]. Studies of reverse engineers, bug hunters, and hackers document offensive-security-style discovery practices, such as exploratory probing, tool chaining, and opportunistic search strategies [S5, S7, S11, S20, S44]. Finally, ecosystem-focused work shows how detection practices are adapted to contexts such as smart contracts and Decentralised Finance (DeFi) [S35, S36], blockchain platforms [S41], embedded systems [S42], and dependency-heavy software supply chains [S46], where adversarial environments, platform constraints, or limited tool coverage shape how practitioners look for SVs.

\paragraph{\textbf{Practices for SV assessment and prioritisation}}
Assessment practices in the included studies usually appear as part of broader workflows rather than as stand-alone activities. Offensive security practitioners [S7] and application security experts [S32] describe how they judge impact and severity when deciding which findings merit follow-up, combining informal heuristics with more structured reasoning about risk. One study examines how practitioners apply CVSS in real cases [S17], revealing ambiguities in interpreting scoring guidelines and variation in how different teams rate similar SVs. Another investigates how cryptographic library developers assess timing-attack vulnerabilities [S56], showing that some risks are perceived as relatively easy to mitigate and are therefore deprioritised despite their potential impact. Assessment is also embedded in artefacts that practitioners use to structure SVM work, for example when they interpret security-relevant technical-debt markers [S1], weigh the implications of security pull requests and automated updates [S45], or use ecosystem and dependency information to judge which components warrant further analysis or remediation [S57]. Code-review studies likewise show reviewers informally assessing the security relevance of changes alongside functional and maintainability concerns [S23]. Recent work on AI-based SVM support explores how AI-based tools and LLMs are used to triage, explain, and prioritise SVs [S49, S50], while also documenting concerns about the reliability of automated judgements and the risk of over-reliance on such support.

\paragraph{\textbf{Practices for SV repair and patching}}
Repair-oriented practices focus on how identified SVs are turned into concrete changes under real-world constraints. Studies on patch management describe how organisations use automation, scheduling strategies, and cross-team coordination to deploy security patches while maintaining service continuity [S9, S39]. Developer-facing work on security tools shows how the presentation of findings, integration into IDEs and build pipelines, and perceived cost–benefit trade-offs influence whether tools are used to support fixing SVs [S12, S29, S31, S37, S55]. Mobile and app-centric studies highlight how security issues are prioritised and remediated alongside feature work in app ecosystems [S33]. Regression-focused work on SVM illustrates how previously fixed SVs can reappear, and identifies practices for detecting and preventing regression SVs in evolving codebases [S52]. In dependency-heavy settings, teams depend on upgrade policies, dependency management routines, and Software Bill of Material (SBOM)-enabled analyses to remediate SVs introduced through third-party components [S46, S54]. Emerging research on AI-based support for repair investigates how AI-based tools and LLMs are used to propose patches, explain fixes, or draft remediation steps [S49, S50, S51], while emphasising the need for human review and validation of machine-generated changes.

\paragraph{\textbf{Practices for SV disclosure}}
Only a small subset of the studies deals explicitly with disclosure and compliance, and these focus on how organisations coordinate SV information rather than on formal disclosure standards. Work on application security support describes how security experts guide development teams in communicating SV-related findings, escalating issues, and mediating between technical and managerial stakeholders during development [S32]. Studies of static-analysis-based security testing and false negatives show how missed issues and tool limitations affect trust in reported findings and shape subsequent conversations about which SVs to track and address [S15]. Research on real-world bug discovery and reporting documents the channels through which SVs are reported from outside the organisation, the motivations of finders, and the practical challenges of handling incoming reports at scale [S20, S44]. Compliance-focused work on IoT product security examines how organisations interpret security certification requirements, negotiate expectations with certification bodies, and align certification activities with their broader SVM efforts [S48]. Finally, studies of automated security pull requests and AI-assisted SVM explore how tool-generated notifications and LLM-based suggestions are triaged, discussed, and acted upon [S45, S50], highlighting communication overheads and the need to integrate automated signals into existing coordination practices.

\begin{center}
\fbox{%
\begin{minipage}{0.96\linewidth}
\textbf{Summary of RQ1.2 (practices).}

\begin{itemize}
    \item 42 practice-oriented studies describe SVM practices for real-world contexts, most often as practitioner-grounded guidelines, recommendations, lessons learned, best practices, or shared experiences.
    
    \item Across the SVM life cycle, reported practices most frequently target detection; fewer studies address assessment and repair, and only a small subset focus explicitly on disclosure-related work such as handling external SV reports.
    
    \item Even within practitioner-centred work, the literature focuses more on how SVs are found and fixed during everyday development than on how SVM is performed as an end-to-end process, which motivates the deeper analysis of practitioner-reported challenges in RQ2.
\end{itemize}
\end{minipage}%
}
\end{center}

\subsection{RQ2}
RQ2 asks what challenges do practitioners report when adopting and using SVM solutions in real-world contexts. To answer this question, we analysed the challenge-related data item (D13) for all 57 primary studies using the thematic analysis procedure described in Section~\ref{Sec:data-analysis}. The final codebook comprises 179 codes, which we organised into 31 subthemes and 5 overarching themes (Table~\ref{tab:rq2-themes}).

\begin{table}[htbp]
\centering
\caption{Overview of themes for practitioner-reported SVM challenges (RQ2).}
\label{tab:rq2-themes}
\resizebox{\textwidth}{!}{%
\begin{tabular}{|
    >{\columncolor{blue!10}}C{0.8cm}|
    >{\columncolor{blue!10}}C{4.8cm}|
    C{2.6cm}|
    C{1.5cm}|
    C{1.8cm}|
    C{7.4cm}|}
\hline
\textbf{ID} & \textbf{Theme} & \textbf{\#Subthemes} & \textbf{\#Codes} & \textbf{\#Studies} & \textbf{Summary description} \\
\hline
T2 & Tool adoption, usability, and integration challenges & 10 & 74 & 42 &
Limitations of security tools and automation in day-to-day development and operations workflows, including hard-to-interpret feedback, fragile integration, poor usability, costly configuration, noisy or ambiguous findings, limited coverage, fragmented security evidence across work systems, and low trust in analysis results. \\
\hline
T1 & Organisational conditions, processes, and capacity constraints & 9 & 51 & 40 &
Organisational structures, processes, culture, governance, and resource limitations that constrain how SVM work is staffed, prioritised, coordinated, and sustained across teams and over time. \\
\hline
T3 & Information, intelligence, and risk-assessment difficulties & 5 & 26 & 15 &
Gaps and misalignments in the information, intelligence, metrics, and standards practitioners rely on to understand SV exposure and justify SVM decisions, including contextual risk information, SV and exploit intelligence, SBOMs, and security-debt markers, which make risk-based triage and planning difficult in practice. \\
\hline
T4 & System architecture, dependency, and domain constraints & 5 & 18 & 13 &
Complex code and system structure, dependency and update management, performance and resource trade-offs, and domain-specific constraints (e.g., embedded systems, smart contracts) that hinder security reasoning, localisation, prioritisation, and remediation of SVs. \\
\hline
T5 & Detection, scanning, and review constraints & 2 & 10 & 7 &
Practical, ethical, and legal limits on detection, scanning, testing, and review activities, including partial or shallow security checks and concerns about intrusive assessment, which leave some systems, changes, or targets under-tested or unprobed. \\
\hline
\end{tabular}%
}
\end{table}

The themes capture recurrent barriers related to organisational conditions and capacity, tool adoption and integration, information and risk assessment, system architecture and dependencies, and constraints on detection, scanning, and review. Two themes are pervasive: \textbf{\textit{Tool adoption, usability, and integration challenges (T2)}} is reported in 42 studies and \textbf{\textit{Organisational conditions, processes, and capacity constraints (T1)}} in 40 studies, while the remaining three themes, \textbf{\textit{Information, intelligence, and risk-assessment difficulties (T3)}}, \textbf{\textit{System architecture, dependency, and domain constraints (T4)}}, and \textbf{\textit{Detection, scanning, and review constraints (T5)}}, appear in 15, 13, and 7 studies, respectively. In the remainder of this subsection, we present the themes in descending order of the number of contributing studies and unpack each theme with its subthemes and illustrative examples from the primary studies.

\subsubsection{\textbf{Tool adoption, usability, and integration challenges (T2)}}
\begin{figure}[htbp]
\centering
\includegraphics[width=0.9\textwidth]{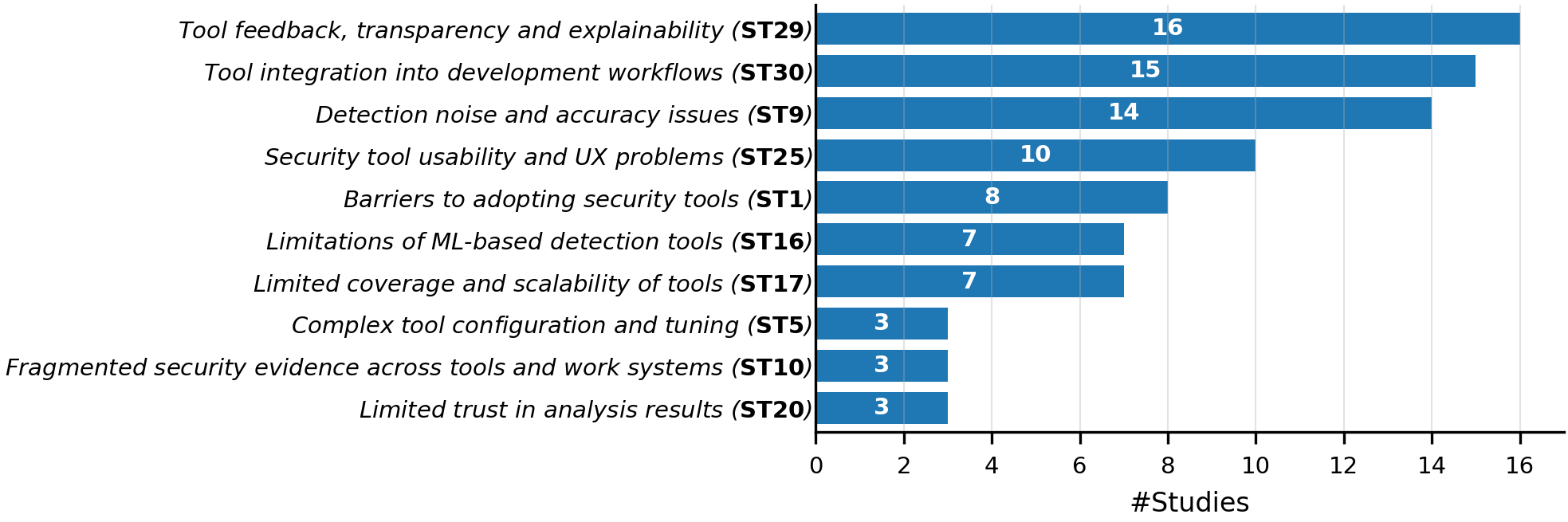}
\caption{Subthemes under Theme T2 (Tool adoption, usability, and integration challenges), by contributing primary studies.}
\label{fig:rq2-t2-subthemes}
\end{figure}

This theme brings together challenges related to adopting, using, and integrating security tooling into day-to-day development and operations workflows. It spans 10 subthemes and 74 codes and is supported by 42 primary studies (Table~\ref{tab:rq2-themes}). As summarised in Figure~\ref{fig:rq2-t2-subthemes}, the most frequently represented subthemes concern tool feedback and explainability, workflow integration, and detection noise and accuracy issues. Overall, the theme shows that even technically capable tools can underperform in practice when feedback is hard to interpret, workflow integration is fragile, usability is poor, configuration is costly, or trust in analysis results is low. The full subtheme descriptions and study mappings are provided in the online appendix \cite{svm-slr-online-appendix}.

A first group of subthemes concerns how tools surface information and fit into everyday work. \textit{Tool feedback, transparency and explainability (ST29)} captures reports that security tools provide limited, poorly structured, or delayed feedback; developers struggle to see what changed, why a finding matters, or how to respond, which undermines their ability to act on results. For example, asynchronous cloud-based SAST runs in the background with little progress indication, so developers are unsure when results are ready or whether they still match the current code state [S3], and static-analysis warnings are often poorly explained or lack actionable context, making it hard to decide whether and how to address them [S29, S31]. AI-assisted tools can also flag issues without clearly justifying their suggestions, which further weakens confidence in their output [S40]. \textit{Tool integration into development workflows (ST30)} reinforces this picture: security tools are often bolted onto IDEs, CI/CD, or patch pipelines in ways that do not match existing practices, so feedback arrives at the wrong time or in the wrong place and is easily ignored. Studies report, for instance, that cloud-hosted SAST is decoupled from local development and code review [S3], SV scanners and patch-orchestration systems are not fully integrated in patch management workflows [S9], security tools mandated by organisational programs sit outside teams’ established issue trackers and pipelines [S28, S55], and AI-based or ML-based security tools are difficult to plug into existing triage and remediation processes [S49]. \textit{Security tool usability and UX problems (ST25)} add that cluttered interfaces, unprioritised result lists, and awkward navigation make it hard to understand and resolve findings in practice: developers perceive security tools as hard to use and too effortful to fit into time-constrained work [S4, S24], empirical studies report that tools with high false-positive rates and awkward workflows discourage regular use [S20], and both traditional and formal-methods-based tools are described as confusing or overly complex, leading developers to disable or avoid them in everyday development [S32, S56].

A second group of subthemes focuses on the reliability and scope of the findings that tools produce. \textit{Detection noise and accuracy issues (ST9)} are widespread: static analysis, SCA, and specialised scanners routinely produce high false-positive rates, miss certain classes of SVs, or report results with insufficient context, forcing teams to invest substantial effort in manual verification and triage. This pattern appears in comparative studies of penetration-testing tools and scanners [S7], in industrial evaluations of static analysis where false positives, limited coverage, and other accuracy issues are seen as major obstacles to adoption [S15, S26, S31], and in smart-contract and program-analysis settings where tools disagree, contradict expert judgement, or miss semantically subtle issues [S10, S36]. Practitioners in dependency-heavy ecosystems describe being flooded with alerts from Software Composition Analysis (SCA) tools while still lacking precise exploitability information, which leads to scepticism and alert fatigue [S42, S45, S46, S57]. \textit{Limitations of ML-based detection tools (ST16)} capture similar concerns for ML- and AI-based tooling: security teams report that ML detectors and LLM-based assistants can misinterpret prompts, over- or under-report SVs, or focus on superficial patterns, so their findings must be treated as tentative and carefully cross-checked [S6, S10, S12, S40, S50]. Opaque decision-making and model uncertainty make it hard for practitioners to understand why a particular SV was flagged (or missed), especially when training data is scarce or biased toward specific languages and frameworks [S12, S40, S43, S49]. \textit{Limited coverage and scalability of tools (ST17)} further emphasises that no single tool covers all relevant SV types or all layers of complex systems; language, framework, and scale constraints leave important areas under-analysed, especially for large, interconnected dependency graphs. Practitioners report, for example, that security tools do not adequately cover protocol logic or certain bug classes [S5, S7], struggle with large or highly optimised codebases [S8, S9], or cannot feasibly explore the full attack surface given time and resource constraints [S27, S31, S32].

Adoption and ongoing configuration challenges form a third cluster. \textit{Barriers to adopting security tools (ST1)} describe how setup costs, learning curves, and uncertainty about value make teams hesitant to commit to new tools. Developers report that asynchronous or cloud-based analysis adds overhead to local workflows and is therefore rarely used [S3, S8], that introducing static analysis requires time for training and experimentation that many teams cannot spare [S15, S24], and that mandated security tools in specialised domains, such as embedded or systems programming, are perceived as ill-suited to local constraints [S42, S44, S53]. In some cases, security tools are introduced top-down as part of organisational security programs, and are perceived primarily as compliance obligations rather than helpful engineering aids, which results in shallow engagement and selective use [S28, S31]. Even when tools are adopted, \textit{Complex tool configuration and tuning (ST5)} remains a source of friction: program analysis tools and patch-automation pipelines require substantial effort to configure and fine-tune, and default settings rarely match organisational risk profiles [S9, S31]. For ML-based tools, practitioners highlight the additional burden of curating training data, managing models, and keeping configurations aligned with evolving systems [S49]. These overheads make it difficult to keep tools “always on” and reduce the practical benefit of automation.

Finally, \textit{Fragmented security evidence across tools and work systems (ST10)} and \textit{Limited trust in analysis results (ST20)} show how evidence fragmentation and trust interact. Security-relevant evidence is scattered across scanner dashboards, bug-bounty portals, issue trackers, repositories, and external advisories, so practitioners lack a single integrated view for detection and triage [S5, S11, S29]. Meanwhile, vague or poorly evidenced warnings, missing reproduction steps, and weak linkage between findings and concrete risks undermine confidence in tool output: studies report skepticism toward static-analysis alerts in industrial settings [S14], caution when relying on automated repair frameworks that may generate incorrect or incomplete fixes [S21], and doubts about the accuracy of AI-based assistants in production codebases [S40]. Practitioners must manually assemble and interpret this dispersed evidence to form a coherent view of SV exposure and decide what to fix first, which delays action and increases the cognitive load of SVM work.

Collectively, \textbf{\textit{Tool adoption, usability, and integration challenges (T2)}} indicates that adoption, trust, and integration challenges are central determinants of whether security tooling meaningfully supports SVM. Tools that produce technically sophisticated analyses can still fail in practice when they are noisy, poorly explained, costly to configure, or misaligned with how teams plan, triage, and implement SV-related work.

\subsubsection{\textbf{Organisational conditions, processes, and capacity constraints (T1)}}
\begin{figure}[htbp]
\centering
\includegraphics[width=0.8\textwidth]{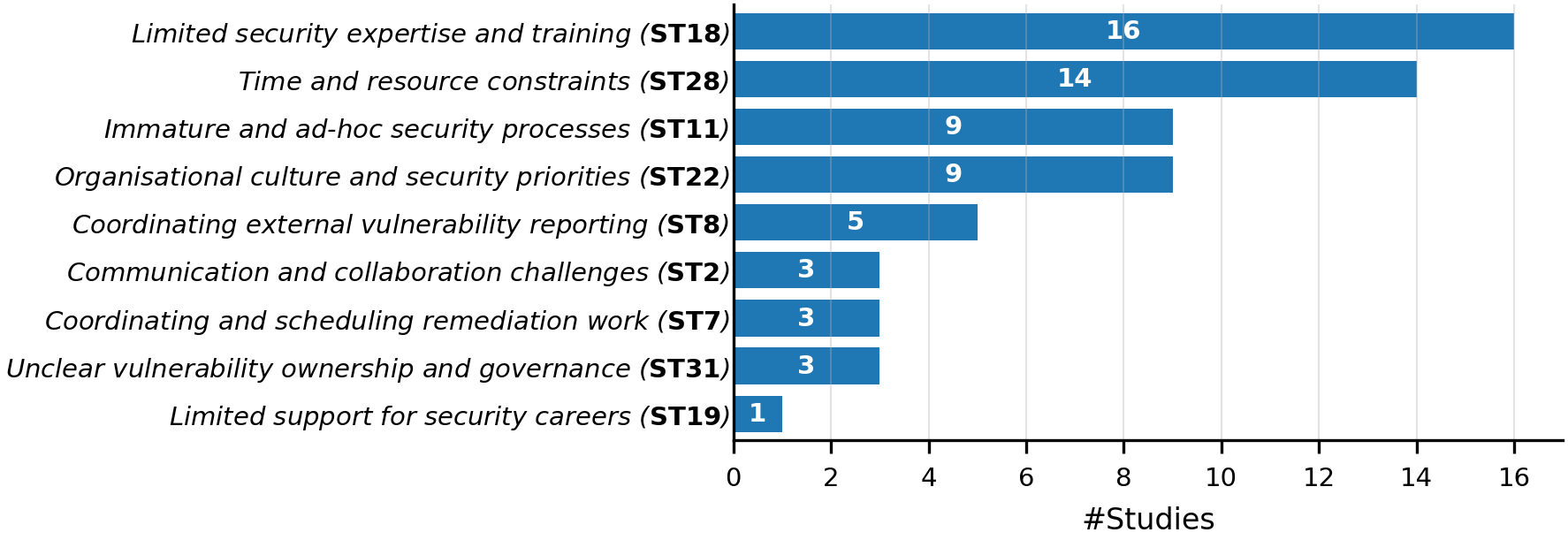}
\caption{Subthemes under Theme T1 (Organisational conditions, processes, and capacity constraints).}
\label{fig:rq2-t1-subthemes}
\end{figure}

This theme organises practitioner-reported challenges around how organisational structures, roles, and resourcing conditions shape SVM work. It spans 9 subthemes and 51 codes and is supported by 40 primary studies (Table~\ref{tab:rq2-themes}). As summarised in Figure~\ref{fig:rq2-t1-subthemes}, the most frequently represented subthemes concern limited security expertise and training, time and resource constraints, and immature or fragmented processes. Overall, the theme primarily concerns limited security expertise and training, constrained time and resources, immature or fragmented processes, and unclear ownership and incentives for SVM activities. The full subtheme descriptions and study mappings are provided in the online appendix \cite{svm-slr-online-appendix}.

A first group of subthemes concerns the skills and capacity that organisations can bring to SVM tasks. \textit{Limited security expertise and training (ST18)} is a recurring concern: developers who are asked to perform security work alongside their usual tasks report that they lack specialised knowledge of secure coding patterns, security APIs, and attack techniques, and often have no access to security specialists for consultation [S4, S22, S23, S33, S51]. Studies on code review and app development show reviewers struggling to identify security-relevant changes or explain security concerns clearly, even when they recognise that something “looks risky” [S23, S33]. Work on SV discovery and bug hunting likewise reports that practitioners find it hard to develop systematic skills for SV discovery without structured training or mentorship [S5, S13]. In several domains, including embedded systems and cryptography, practitioners highlight the lack of training materials and examples that show how to avoid or repair SVs in their particular stacks [S25, S42, S56]. Complementing this, \textit{Limited support for security careers (ST19)} notes that SV discovery and related security work can have uncertain economic and career prospects, which discourages sustained investment in building deep expertise [S13].

\textit{Time and resource constraints (ST28)} reinforce these capability gaps. Across studies, developers and managers describe pressure to deliver features and meet deadlines, leaving little room for in-depth security analysis, systematic scanning, or careful patching [S22, S23, S24, S37, S47]. Security activities such as code review for SVs, dependency updates, and patch testing are often squeezed into remaining time after feature work, or postponed when release dates are tight [S2, S29, S51, S55]. In large organisations, teams report that they cannot scale patching and follow-up testing to keep pace with SV notifications, even when they recognise the importance of timely remediation [S45, S46, S56]. These shortages in time and staffing mean that SVM work is often selective and risk-based in practice, rather than comprehensive.

A second group of subthemes focuses on how SVM is organised and fits within organisational culture. \textit{Immature and ad-hoc security processes (ST11)} describes situations where SV-related activities are handled through informal routines, custom scripts, or one-off initiatives rather than stable, institutionalised processes. For example, studies of security metrics and patch management report that SV data is collected and analysed in project-specific ways, with limited standardisation and weak feedback loops into future planning [S18, S47, S55]. In some contexts, security policies and certification requirements exist on paper, but are loosely connected to everyday development and operations, so teams revert to ad-hoc workarounds under pressure [S20, S36, S48, S51]. Closely related, \textit{Organisational culture and security priorities (ST22)} highlights that security and SVM tasks often compete with business and feature priorities. Developers in several studies report that security is framed as “nice to have” or a compliance obligation, making it hard to allocate time for SV scanning, code review, and patching when new features or performance improvements are prioritised [S4, S28, S31, S37]. In fast-moving domains such as blockchain and DeFi, strong economic incentives and pressure for rapid feature delivery similarly lead teams to cut back on thorough audits and patching of known SVs [S35, S41].

The remaining subthemes in \textbf{\textit{Organisational conditions, processes, and capacity constraints (T1)}} concern coordination, governance, and external interfaces. \textit{Communication and collaboration challenges (ST2)} capture how organisational separation between security specialists and development teams, together with unclear channels for discussing findings, leads to misunderstandings about the severity, urgency, or feasibility of fixing SVs [S20, S32]. Application security experts report difficulty in getting development teams to act on findings when communication is routed through formal ticketing systems or infrequent meetings, and developers describe security advice that arrives too late or without sufficient context to be actionable [S20, S32]. \textit{Coordinating external vulnerability reporting (ST8)} adds that relationships with external reporters, bug-bounty participants, and auditors can be strained by slow or opaque responses, unclear expectations, and limited organisational support for handling incoming reports; this can erode trust between reporters and maintainers and complicate coordinated disclosure [S6, S10, S11, S20, S44].

Several subthemes focus on planning and governance of remediation work. \textit{Coordinating and scheduling remediation work (ST7)} shows that, even when fixes are known, arranging patch deployment across teams, services, and environments is difficult because of incomplete asset inventories, tightly coupled systems, and downtime constraints [S4, S9, S39]. Large enterprises struggle to reconcile business continuity requirements and regulatory patching deadlines, which can lead to prolonged exposure to known SVs [S9, S39]. \textit{Unclear vulnerability ownership and governance (ST31)} highlights uncertainty about who is responsible for SV triage and remediation, and ultimately accountable for security decisions: security engineers emphasise that humans must retain authority to approve or reject automated repairs, while large organisations report ambiguous ownership for systems or SV classes, slowing coordinated responses [S21, S39, S51].

Overall, \textbf{\textit{Organisational conditions, processes, and capacity constraints (T1)}} shows that many of the most persistent SVM challenges arise not from purely technical limitations, but from how security work is staffed, prioritised, and governed. Limited expertise and training, chronic time and resource constraints, immature processes, weak security culture, and unclear ownership together constrain what can be achieved with existing SVM approaches and tools, and help explain why known SVs can remain unaddressed even when detection capabilities are available.

\subsubsection{\textbf{Information, intelligence, and risk-assessment difficulties (T3)}}
\begin{figure}[htbp]
\centering
\includegraphics[width=0.85\textwidth]{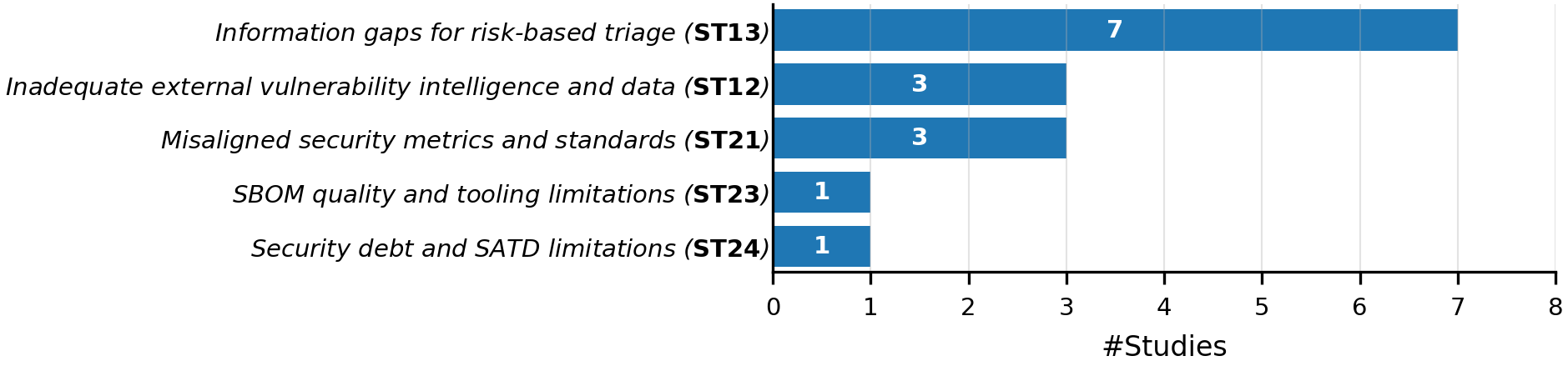}
\caption{Subthemes under Theme T3 (Information, intelligence, and risk-assessment difficulties).}
\label{fig:rq2-t3-subthemes}
\end{figure}

This theme organises practitioner-reported challenges around the information, metrics, and standards that practitioners rely on to understand SV exposure and justify SVM decisions. It spans 5 subthemes and 26 codes and is supported by 15 primary studies (Table~\ref{tab:rq2-themes}). As summarised in Figure~\ref{fig:rq2-t3-subthemes}, the most frequently represented subthemes concern information gaps for risk-based triage, limitations of SV and exploit intelligence, and misaligned security metrics and standards. Overall, the theme concerns gaps in contextual information for risk-based triage, limitations of SV and exploit intelligence, misaligned security metrics and certification requirements, and difficulties operationalising SVM-relevant artefacts such as SBOMs and security-related technical debt. The full subtheme descriptions and study mappings are provided in the online appendix \cite{svm-slr-online-appendix}.

A first group of subthemes concerns the information practitioners need to prioritise which findings matter most. \textit{Information gaps for risk-based triage (ST13)} shows that, even when detection works, teams often lack the contextual data required for confident prioritisation. Penetration testers report that high setup overheads and compounding interpretations during assessment make it difficult to carry out focused testing and to provide clear triage and remediation advice~[S7]. In code review, developers with limited security expertise find it difficult to assess the impact and severity of security-relevant changes, and generic checklists are often too abstract to guide triage decisions~[S34]. Studies on kernel error behaviour and dependency management similarly describe limited visibility into how low-level bugs or SCA alerts translate into real exploitability and operational risk, forcing engineers to invest substantial manual effort in tracing consequences and deciding what to fix first~[S38, S45, S57]. Work on severity scoring further shows that ambiguity in CVSS interpretation leads to inconsistent scores across and even within assessors, which undermines triage decisions that depend on these ratings~[S17]. Recent evaluations of LLM-based support also find that text-focused analysis tends to latch onto superficial patterns rather than full code and system context, limiting its ability to perform reliable exploitability assessment or patch validation and requiring careful manual verification~[S50].

\textit{Inadequate external vulnerability intelligence and data (ST12)} highlights quality and timeliness issues in the external intelligence that SVM decisions depend on. Work on silent fixes reports that available SV and patch information is sparse and heterogeneous, and that there can be long delays between when a fix is implemented and when an advisory or detailed explanation becomes available~[S10]. Patch-management studies similarly describe poorly structured vulnerability information and inadequate vendor documentation, so that automated scanners and analyses still require substantial manual verification before teams can safely act on their results~[S39]. Work on Bayesian exploitability modelling adds that detailed SV intelligence is expensive and slow to acquire, and that many organisations lack the specialised data and expertise needed to deploy such advanced analysis methods in everyday risk assessment~[S30].

A second group of subthemes focuses on how formal metrics and standards relate to decision-making. \textit{Misaligned security metrics and standards (ST21)} shows that the measures and expectations used in practice do not always align with the risks and trade-offs practitioners face. One study reports that teams lack consistent, comparable metrics to track SVM progress or benchmark their posture, and that many organisations adopt security practices mainly to satisfy audits and compliance requirements rather than improve defect prevention and sustained security outcomes~[S18]. Work on attack surface analysis illustrates that, even when quantitative metrics are available, engineers struggle to choose parameters and interpret results in ways that balance attack-surface reduction against development time and resource constraints~[S27]. In regulated domains such as IoT product development, participants describe security certification as an expensive, time-consuming hurdle whose checks are weakly connected to their day-to-day patching and remediation processes, which further limits the perceived value of these standards for guiding SVM work~[S48].

The remaining subthemes capture challenges with specialised artefacts for representing SVM-relevant information. \textit{SBOM quality and tooling limitations (ST23)} synthesises evidence that SBOM-based SVM remains difficult in practice: business stakeholders lack expertise to interpret SBOM entries, current SBOMs are often incomplete or too coarse to support concrete decisions, producing and maintaining them demands substantial time and effort, misclassified or noisy SBOM entries require manual triage, and immature tooling and competing standards make it hard to integrate SBOMs into build and release processes~[S54]. \textit{Security debt and SATD limitations (ST24)} shows that encoding security concerns as Self-Admitted Technical Debt (SATD) is a double-edged practice: while SATD comments are intended to record outstanding security obligations, they can reveal sensitive information to attackers, embed low-quality or misleading security guidance, and complicate SV reporting and coordination when they spread across code and issue trackers~[S1].

In summary, \textbf{\textit{Information, intelligence, and risk-assessment difficulties (T3)}} underscores that risk-based SVM depends on information and metrics that are often incomplete, delayed, costly to obtain, or difficult to interpret in context. Even when detection and repair capabilities are available, practitioners face challenges in assembling trustworthy, decision-ready views of SV exposure and grounding SVM decisions on transparent, shared understandings of risk.

\subsubsection{\textbf{System architecture, dependency, and domain constraints (T4)}}
\begin{figure}[htbp]
\centering
\includegraphics[width=0.85\textwidth]{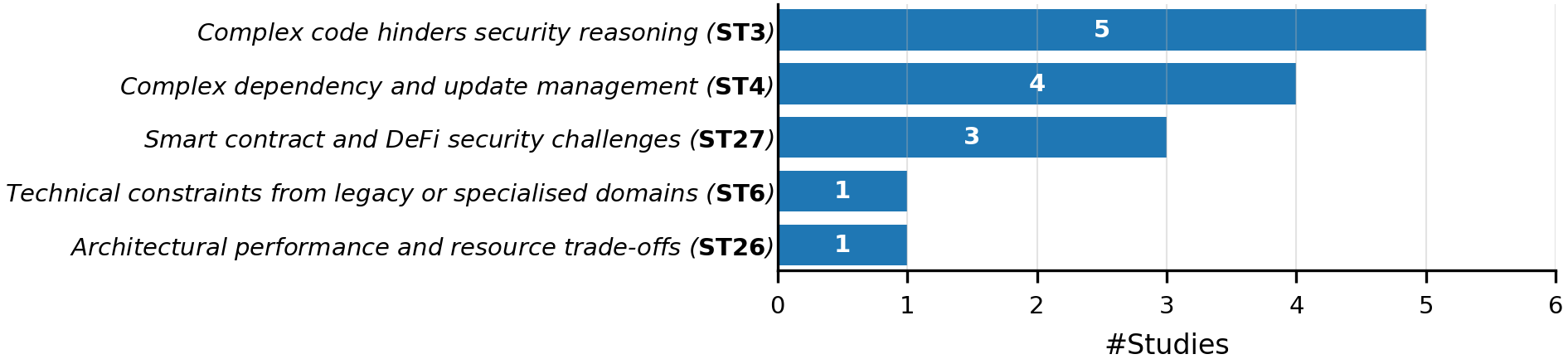}
\caption{Subthemes under Theme T4 (System architecture, dependency, and domain constraints).}
\label{fig:rq2-t4-subthemes}
\end{figure}

This theme synthesises how system structure, dependencies, and domain constraints shape SVM feasibility. It spans 5 subthemes and 18 codes and is supported by 13 primary studies (Table~\ref{tab:rq2-themes}). As summarised in Figure~\ref{fig:rq2-t4-subthemes}, the most frequently represented subthemes concern complex code and architectures, dependency and update management, and domain-specific challenges in settings such as smart contracts and DeFi. Overall, the subthemes show that complex code and architectures hinder security reasoning, dependency-heavy systems introduce persistent remediation and update difficulties, and domain-specific contexts such as smart contracts, embedded systems, and plug-and-play architectures impose constraints on SVM decisions. The full subtheme descriptions and study mappings are provided in the online appendix \cite{svm-slr-online-appendix}.

A first group of subthemes concerns how code and system structure affect practitioners’ ability to reason about security. \textit{Complex code hinders security reasoning (ST3)} captures reports that intricate or hard-to-understand designs make it difficult for developers and security experts to trace security-relevant behaviours, anticipate side effects, and safely remediate issues. For example, Rust developers working with \texttt{unsafe} blocks describe how the complexity of understanding and maintaining these regions, combined with tooling that only analyses local crates, creates gaps between analysis scope and real SV exposure [S8]. Android developers report that implementing secure dynamic class loading correctly requires navigating cryptographic details and certificate validation logic that many find difficult to reason about, especially when migrating existing apps to more secure frameworks [S25]. Kernel engineers note that limited visibility into diverse error behaviours and the effort needed to trace corrupted state across the kernel make it hard to assess exploitability and prioritise bug fixing [S38]. Similar concerns appear in large-system contexts, where complex code paths make it hard to see how changes might introduce new SVs or regressions [S52].

A second group of subthemes highlights challenges of working in dependency-heavy ecosystems. \textit{Complex dependency and update management (ST4)} shows that managing SVs in large dependency graphs is difficult in practice. Reviewers report that heavy reliance on third-party libraries leaves blind spots in code review because they cannot see or evaluate library internals, even when they suspect security-relevant effects [S23]. In ecosystems with automated dependency updates and SCA tooling, developers describe being overwhelmed by alerts and security pull requests, struggling to identify which updates matter, and facing substantial review and testing overhead for each change [S45, S46]. Teams also report that fear of breaking changes, build failures, or dependency conflicts leads them to defer or avoid applying security updates, creating long-lived dependency-related security debt despite known upstream SVs [S46, S57].

\textit{Smart contract and DeFi security challenges (ST27)} capture domain-specific architectural constraints in blockchain systems. Even security-aware practitioners report that, despite using analysis tools and reviews, SVs still reach deployed contracts because formalised processes are weak or absent [S35]. DeFi projects face pressure from rapid release cycles and strong economic incentives, which can push teams to prioritise shipping functionality over thorough security testing and auditing [S35, S41]. Practitioners also note that current tools rarely cover logic and protocol-layer vulnerabilities in DeFi systems, forcing them to rely on manual reasoning and expert audits for the most critical issues [S36].

The remaining subthemes emphasise how legacy environments and performance constraints limit remediation options. \textit{Technical constraints from legacy or specialised domains (ST6)} show that in embedded and low-level contexts, immature analysis and verification tools, interoperability requirements with legacy C code, and limited hardware and toolchain support make it difficult to use memory-safe languages such as Rust as a reliable boundary [S42]. \textit{Architectural performance and resource trade-offs (ST26)} highlight that recommended mitigation patterns can impose unacceptable performance overheads in latency-sensitive architectures; practitioners report that the architectural changes and runtime costs required for stronger controls in plug-and-play systems can exceed resources or degrade user experience, making stronger mitigations difficult to deploy in practice without unacceptable performance or usability costs [S2].

Taken together, \textbf{\textit{System architecture, dependency, and domain constraints (T4)}} underscores that many challenges are rooted in the structural properties of the systems and ecosystems studied. Complex codebases, deep dependency stacks, and domain-specific constraints limit which detection, assessment, and repair strategies are realistic in practice, and help explain why some SV-related risks persist even in teams actively trying to improve their SVM practices.

\subsubsection{\textbf{Detection, scanning, and review constraints (T5)}}
\begin{figure}[htbp]
\centering
\includegraphics[width=0.8\textwidth]{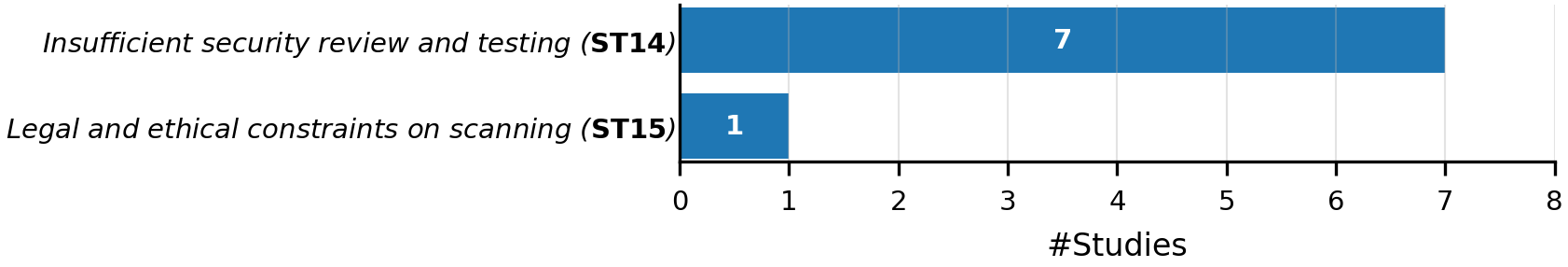}
\caption{Subthemes under Theme T5 (Detection, scanning, and review constraints).}
\label{fig:rq2-t5-subthemes}
\end{figure}

This theme synthesises constraints that limit detection, scanning, and security review in practice. It spans 2 subthemes and 10 codes and is supported by 7 primary studies (Table~\ref{tab:rq2-themes}). As summarised in Figure~\ref{fig:rq2-t5-subthemes}, the two subthemes indicate that security reviews and tests are often partial or constrained and that legal and ethical concerns can restrict large-scale scanning, even when practitioners are aware of the techniques. The full subtheme descriptions and study mappings are provided in the online appendix \cite{svm-slr-online-appendix}.

\textit{Insufficient security review and testing (ST14)} captures ways security checks fall short in day-to-day work. Offensive security practitioners report that tight engagement timelines and tool setup overheads constrain how many targets and paths can be examined in depth, forcing them to prioritise a subset of targets and leaving parts of the system under-tested [S7]. Studies of code review show that reviewers tend to focus on functional correctness, with security having low importance; many reviewers lack the security knowledge to identify subtle issues and assume that separate assessments will catch them, resulting in shallow or inconsistent security review coverage [S23, S34]. In dependency-heavy contexts, teams describe difficulty thoroughly reviewing and testing all incoming security- or update-related changes: the volume of alerts and security pull requests, combined with the cost of regression testing, leads to selective testing and long-lived unassessed updates [S45]. IoT product teams note that formal certification and conformance testing can encourage a narrow focus on passing prescribed checklists rather than on building robust, ongoing SV detection and remediation practices into development and maintenance [S48]. Finally, studies of bug fixing report that existing test suites provide limited coverage of security-relevant scenarios; developers often rely on functional regression tests and over-trust automated tools, so security regressions introduced during bug fixing frequently go unnoticed [S52]. In combination, these findings show that security reviews and tests are frequently partial, time-bounded, or misaligned with actual risk, allowing important issues to slip through [S7, S23, S34, S45, S48, S52].

\textit{Legal and ethical constraints on scanning (ST15)} highlight that practitioners cannot always apply detection techniques as aggressively as they would like. Operators of large-scale server-side scanning report legal uncertainty: even benevolent scans to improve security can be interpreted as unauthorised access or interference, and possible civil liability for perceived harm discourages broad use [S16]. They also worry that aggressive probing can overload or crash infrastructure and impose bandwidth and resource costs on server operators without clear benefit, creating friction and resistance to large-scale scanning campaigns [S16]. At Internet scale, obtaining explicit consent for scanning is infeasible, so teams struggle to reconcile ethical expectations of consent with the technical realities of SV discovery [S16]. As a result, some forms of large-scale active detection remain underused, despite their potential to surface SVs in deployments.

Overall, \textbf{\textit{Detection, scanning, and review constraints (T5)}} shows that practical limits on the depth, scope, and permissibility of review, testing, and scanning leave gaps in SVM coverage. Even when detection techniques and tools are available, these constraints mean that some SV classes and system parts receive limited or inconsistent scrutiny.


\section{Discussion, implications, and future directions for SVM research and practice}
\label{sec:discussion}
\subsection{Cross-cutting synthesis of the SVM landscape}
\label{sec:discussion-synthesis}

Taken together, the results for RQ1 and RQ2 portray an SVM landscape that is technically active and strongly shaped by organisational and contextual constraints. Across the 57 primary studies, practitioner-centred work on SVM spans a range of domains and methods, yet the evidence base is unevenly distributed. Most solutions and investigations concentrate on SV detection and SV repair, while other phases and stakeholder perspectives receive limited attention.

From a solution perspective (RQ1), the corpus is dominated by practice-oriented contributions. Most studies report \emph{practices} such as guidelines, recommendations, lessons learned, best practices, and shared experiences, with a smaller but still substantial subset introducing or evaluating concrete \emph{approaches and tools}. When technical solutions are proposed with practitioner involvement, they are often implemented tools or methods embedded in specific workflows, with fewer abstract frameworks, algorithms, or techniques. In terms of SVM phases, SV detection and SV repair account for most of the reported solutions and evaluations, whereas SV assessment and SV disclosure are addressed by fewer studies and are often treated as secondary concerns within broader detection- or patching-focused work. This imbalance suggests that, in current practitioner-facing research, SVM is still primarily framed as a problem of finding and fixing SVs, rather than as a full life cycle that also encompasses nuanced risk assessment and coordinated disclosure.

The thematic analysis of practitioner-reported challenges (RQ2) adds a complementary socio-technical view. The five themes (T1--T5) show that difficulties with SVM arise not only from limitations of individual tools or techniques, but from how those tools and practices are situated in organisations and systems. \textbf{\textit{Organisational conditions, processes, and capacity constraints (T1)}} and \textbf{\textit{Tool adoption, usability, and integration challenges (T2)}} together indicate that barriers accumulate both around \emph{who} does SVM work and around \emph{how} tools behave in that work: organisations report uneven and scarce security expertise, chronic time and resource pressure, immature or ad-hoc processes, weak security priorities, and unclear ownership and coordination, while the tools they deploy can be noisy, hard to interpret, limited in coverage, difficult to configure, only loosely integrated into everyday workflows, and associated with fragmented security evidence across tools and work systems, which undermines trust and sustained use. \textbf{\textit{Information, intelligence, and risk-assessment difficulties (T3)}} shows that the information infrastructure around SVM, including external SV, exploit, and patch intelligence, severity metrics, SBOMs, and security-debt artefacts, is often incomplete, inconsistent, delayed, or hard to interpret, so teams struggle to assemble shared, decision-ready views of risk for triage and planning. \textbf{\textit{System architecture, dependency, and domain constraints (T4)}} highlights that complex codebases, deep and fragile dependency stacks, domain-specific constraints (e.g., smart contracts, embedded systems, and legacy platforms), and architectural performance and resource trade-offs impose practical limits on which detection, assessment, and remediation strategies are viable in practice. Finally, \textbf{\textit{Detection, scanning, and review constraints (T5)}} underscores that practical limits on the depth, scope, and permissibility of review, testing, and large-scale scanning mean that security scrutiny is frequently partial or selectively applied, leaving systematic gaps in SVM coverage.

Overlaying the RQ1 solutions with these RQ2 themes highlights several cross-cutting patterns. First, the areas where research effort is concentrated do not always align with the bottlenecks that practitioners report. Tool- and detection-focused solutions for SV detection and SV repair are plentiful, yet T1 and T2 show that many teams struggle to secure stable capacity for SVM work and to keep tools embedded and trusted in day-to-day workflows. Second, T3 to T5 help explain why SV assessment and SV disclosure remain less developed in practitioner-centred work: the information needed for risk-based triage is patchy and hard to combine, architectural and ecosystem constraints limit what can realistically be tested or fixed, and legal and ethical considerations restrict some forms of large-scale probing. Taken together, these patterns suggest that improving SVM in practice is at least as dependent on strengthening organisational, informational, and architectural preconditions as it is on introducing new detectors or repair mechanisms.

The empirical foundation of this picture is skewed toward certain perspectives and methods. Most studies in the corpus recruit developers and security specialists as participants, with fewer contributions capturing the views of managers, operators, auditors, or end users. Questionnaires and interviews are the dominant data-collection methods, with relatively few in-situ observations, controlled experiments, or longitudinal field studies. As a result, current practitioner-centred evidence for SVM is rich in self-reported perceptions, experiences, and preferences, but provides less insight into long-term behavioural change or organisational outcomes when new tools or practices are introduced.

Overall, the synthesis suggests that practitioner-centred SVM research has begun to map both the solution space and the challenge space, but the two are not well aligned. Tool- and detection-focused solutions coexist with persistent socio-technical obstacles around capacity, coordination, information, and system constraints. In the following subsections, we examine how our review complements prior SVM-related SLRs, analyse where current solutions do and do not address reported challenges, and outline implications and future directions for SVM research, practice, and tool building.

\subsection{Positioning our review relative to prior SVM and security reviews}
\label{sec:discussion-positioning}

Section~\ref{Sec:Background} has already compared our scope against existing SVM-related surveys and SLRs along three axes: whether they adopt a practitioner-centred lens, which SVM tasks they target (SV detection (D), SV assessment (A), SV repair (R), and SV disclosure (Di)), and their main contribution. Building on that mapping, we now explain how our results advance and refine the picture provided by these prior reviews.

First, our findings add practitioner evidence to detection-focused reviews. Existing work on ML- and DL-based SV detection consolidates algorithms, code representations, datasets, and evaluation metrics, and repeatedly notes dataset quality and benchmarking issues as open problems \cite{ghaffarian2017software,croft2022data,sonnekalb2022deep,bassi2023systematic,kalouptsoglou2023software,senanayake2023android,shiri2024systematic,shimmi2025ai,wang2025review}. Our synthesis shows how these technical limitations manifest in use. Theme \textbf{\textit{T2}} (Tool adoption, usability, and integration challenges) and theme \textit{\textbf{T3}} (Information, intelligence, and risk-assessment difficulties) document that practitioners experience high false-positive rates, inconsistent coverage across languages and stacks, opaque ML- or AI-based judgements, fragmented security evidence across tools and work systems, and incomplete or delayed SV and exploit intelligence as barriers to trusting and acting on detection results. These patterns are visible not only for learning-based tools but also for more traditional static analysis, SCA, and program-analysis tooling. In this way, our RQ2 results turn the dataset and evaluation concerns raised in detection reviews into concrete accounts of alert fatigue, verification overhead, and scepticism about automated findings, showing why many detection advances have only partial impact in industrial SVM workflows.

Second, our cross-cutting socio-technical themes extend repair- and patch-management-focused reviews beyond a single phase. Dissanayake et al.\ \cite{dissanayake2022software} identify socio-technical challenges in patch management and emphasise limited industrial evaluation of proposed approaches. Lin et al.\ \cite{lin2024vulnerabilities} and Fu et al.\ \cite{fu2025ai} trace how SV and patch information flows through OSS ecosystems and DevSecOps pipelines, while Zhou et al.\ \cite{zhou2025large} highlight that LLM-based repair is mostly assessed offline. Our results show that similar organisational and integration problems arise across multiple parts of the SVM life cycle. Themes \textit{\textbf{T1}} (Organisational conditions, processes, and capacity constraints) and \textbf{\textit{T2}} (Tool adoption, usability, and integration challenges) together indicate that limited security expertise and training, time and resource pressure, immature or ad-hoc processes, weak security priorities, unclear ownership, noisy or poorly explained findings, and fragile workflow integration are reported around detection and repair work and also appear, to a lesser extent, in assessment and disclosure-related activities. At the same time, our RQ1 mapping indicates that most approach or tool papers are evaluated in short-term, small-scale studies, with little evidence about sustained industrial use, corroborating evaluation gaps noted in patch-management and AI-repair reviews. Together, these findings move beyond listing patch-management challenges and suggest that similar socio-technical obstacles constrain the adoption and effectiveness of SVM solutions across phases, regardless of whether those solutions are AI-based or not.

Third, by spanning all four SVM tasks with a practitioner-centred lens, our review connects areas that prior secondary work has often treated separately. Existing reviews typically focus on one task or a narrow subset, such as detection \cite{ghaffarian2017software,sonnekalb2022deep,shiri2024systematic,shimmi2025ai,wang2025review}, data-driven assessment and exploitability estimation \cite{le2022survey,elder2024survey}, or patch management and repair \cite{dissanayake2022software,lin2024vulnerabilities,fu2025ai,zhou2025large}. Our RQ1.1 mapping shows that, when practitioners are involved, evaluated approaches and tools are heavily skewed toward repair, often combined with detection, while assessment is covered in fewer studies, and only one practitioner-evaluated tool explicitly targets disclosure. Our RQ1.2 synthesis further shows that practice-oriented guidance concentrates on detection, with repair addressed less often and only a small number of studies focus explicitly on disclosure work such as handling external SV reports. RQ2 then adds that information and risk-assessment problems (T3) and architectural and dependency constraints (T4) shape how teams prioritise and act on SVs across these phases, and that constraints on detection, scanning, and review (T5) mean that some systems and SV classes receive only partial or episodic scrutiny even when relevant techniques are known. This cross-phase, practitioner-grounded view advances the existing body of reviews by turning task-specific literatures into an integrated SVM landscape where tool-centric advances, organisational conditions, and information infrastructure can be analysed together, and where underserved areas such as SV assessment and SV disclosure emerge clearly as priorities for future work.

\subsection{Alignment and gaps between solutions (RQ1) and challenges (RQ2)}
\label{sec:alignment-gaps}

In this subsection, we discuss how the types of solutions identified in RQ1 relate to the practitioner-reported challenges synthesised in RQ2. Our goal is not to claim that the same primary studies systematically evaluated each solution against all challenges. Instead, we use the RQ2 themes as a lens on the RQ1 solution space: for each theme, we summarise which approaches, tools, and practices have been proposed or evaluated in settings where that type of challenge arises, and where practitioners still report that these challenges remain weakly addressed.

For \textbf{T1 (Organisational conditions, processes, and capacity constraints)}, there is partial alignment with the practice-oriented solutions in RQ1, but many challenges remain loosely addressed. Several studies report that developers asked to carry out SV work alongside their usual tasks lack specialised security training and access to security specialists, and are often unsure about the capabilities and limitations of the tools they use [S3, S4, S22, S23]. Code review and app-development studies show reviewers who sense that changes are risky but struggle to articulate or justify those concerns clearly [S23, S33, S34]. Time and resource constraints compound these skills gaps: teams describe pressure to prioritise feature work, limiting capacity for scanning, triage, and careful patching [S22, S23, S24, S37, S51, S56]. At the same time, SVM processes are often organised through ad-hoc scripts or one-off initiatives, and security policies or certification requirements exist on paper but are weakly connected to day-to-day practice [S18, S20, S28, S48, S51, S55].

RQ1 identifies guidelines, recommendations, and other practices that speak to this picture, such as guidance for secure code review and broader security-focused development practices [S23, S34, S51], recommendations for integrating SV considerations into secure development and DevSecOps-style workflows [S28, S37], and suggestions for improving communication and coordination around SV reporting and handling, between external reporters and maintainers and between internal security and development teams [S6, S10, S20, S44]. However, few contributions describe organisational interventions that change staffing models, training programmes, incentive structures, or vulnerability ownership at scale. Large enterprises report difficulties coordinating remediation windows and clarifying responsibility for SV triage and repair, even when patch-management processes and tools are in place [S4, S9, S21, S39]. As a result, many of the most frequently reported T1 challenges appear in RQ1 as motivating context rather than explicit solution targets.

For \textbf{T2 (Tool adoption, usability, and integration challenges)}, alignment is almost inverted. RQ1 shows a rich set of studies that introduce or investigate solutions for SV detection and SV repair with practitioners, often using task-based or mixed-method evaluations [S2, S3, S7, S12, S14, S15, S21, S24, S26, S29, S31, S35, S40, S43, S49, S50, S53, S56]. At the same time, the RQ2 themes indicate that adoption, usability, and workflow integration remain central pain points. Developers report that asynchronous or cloud-based analysis runs in the background with little progress indication, so they are unsure when results are ready or whether they still match the current code state [S3]. Static-analysis tools often provide warnings that are poorly explained or lack actionable context, making it hard to decide whether and how to respond [S29, S31, S32]. Integration with IDEs, CI/CD, and review practices is frequently fragile or missing, so feedback arrives at the wrong time or in the wrong place and is easily ignored [S3, S9, S14, S28, S34, S52, S55, S56]. 

Usability studies add that cluttered interfaces, unprioritised result lists, and awkward workflows discourage regular use of security tools [S4, S7, S20, S24, S32, S56]. ML-based and AI-based tools raise concerns about opacity and instability, since practitioners cannot easily understand why particular SVs are flagged or missed and worry about hallucinations or biased training data [S12, S40, S43, S49, S50]. Configuration and tuning require substantial effort, especially for program-analysis pipelines and ML-based detectors, and default settings rarely align with organisational risk profiles [S9, S31, S49]. Some RQ1 approaches and tools explicitly aim to reduce friction or provide more actionable feedback, but many evaluations emphasise technical performance, such as accuracy or recall, rather than sustained use in real workflows. The persistence of T2 challenges suggests a gap between the problems tools are designed to solve and the problems practitioners encounter when trying to keep those tools active and trusted in day-to-day SVM work.

\textbf{T3 (Information, intelligence, and risk-assessment difficulties)} sits at the intersection of RQ1 contributions on metrics, risk models, SBOMs, and security-debt artefacts and the RQ2 findings about the difficulty of assembling decision-quality SV information. Several RQ1 studies introduce quantitative security metrics and risk models, including attack-surface measures and exploitability models [S27, S30, S47], alongside empirical work on SV severity assessment and CVSS interpretation [S17, S30], and SBOM-based workflows or SATD representations intended to make SV-relevant information more explicit [S1, S54]. However, RQ2 shows that teams often lack the contextual information needed for risk-based triage, even when detection works. Penetration testers report that heavy setup overhead and compounding interpretations make it difficult to provide focused triage and remediation guidance [S7]. Developers and reviewers describe difficulty assessing the impact and severity of security-relevant changes and note that generic checklists or ambiguous CVSS scores are poor guides to triage [S17, S34]. Engineers working with kernels and dependency-heavy systems report limited visibility into how low-level bugs or SCA alerts translate into real exploitability and operational risk, which forces substantial manual effort to trace consequences [S38, S45, S57]. 

Quality and timeliness issues in external intelligence further complicate planning. Studies on silent fixes and patch management observe that SV information is often sparse, heterogeneous, and delayed relative to when fixes are implemented [S10, S39], while work on Bayesian exploitability modelling emphasises that detailed intelligence is expensive and slow to acquire and requires specialised expertise [S30]. Practitioners also report that SBOMs can be incomplete, noisy, or poorly integrated with existing build and release processes, which limits their value for impact analysis [S54]. In other words, there are solutions that target information and risk assessment, but they are often evaluated offline or in limited settings and rarely followed through to show how they change triage decisions, patch scheduling, or backlog reduction in practice. Many T3 challenges concern how information is interpreted, combined, and trusted across teams, whereas most RQ1 contributions still focus on producing new indicators or artefacts.

For \textbf{T4 (System architecture, dependency, and domain constraints)}, the RQ1 solution space includes domain-specific solutions for systems programming [S2, S8, S42, S56], dependency and supply-chain security [S45, S46, S54, S57], and smart contracts or DeFi [S35, S36, S41]. These contributions address aspects of architectural complexity and domain constraints, for example by analysing unsafe Rust use [S8], providing support for dependency updates [S45, S46], or developing contract-analysis tools for blockchain ecosystems [S35, S36, S41]. Nevertheless, the RQ2 synthesis shows that complex, tightly coupled codebases make it difficult to reason about security and predict the impact of changes [S8, S25, S38, S52]. Deep dependency graphs and transitive relationships complicate update planning and leave teams with long-lived dependency-related SV debt [S23, S45, S46, S57]. Domain-specific environments, such as DeFi protocols and embedded or low-level systems, impose technical constraints related to immutability, composability, legacy interoperability, hardware, toolchain support, and architectural performance or resource trade-offs that limit what SVM strategies are feasible [S2, S35, S36, S41, S42].

Only a minority of RQ1 solutions treat architectural and domain constraints as first-class design concerns, for example by incorporating real-time requirements, legacy interoperability, or composability risks into threat and solution models. Many tools are evaluated on single projects or simplified settings and do not address coordination across heterogeneous stacks or long-lived dependency chains. Consequently, T4 challenges remain only partially covered, especially where architectural and domain constraints interact with organisational and informational issues from T1 and T3.

\textbf{T5 (Detection, scanning, and review constraints)} reveals a gap between what is technically possible and what practitioners feel able to do in real environments. RQ1 documents numerous detection and scanning tools and practices for security review, testing, and measurement, including penetration-testing tools, static and dynamic analysis, SCA, and automated or semi-automated repair support [S1, S2, S3, S6, S7, S10, S11, S12, S14, S15, S16, S18, S20, S23, S24, S26, S32, S36, S38, S40, S42, S45, S46, S48, S50, S52, S57]. These solutions can increase SV coverage. Yet RQ2 shows that security review and testing often remain partial, selective, or shallow in practice. Offensive security practitioners describe tight engagement timelines and tool-setup overhead that force them to prioritise targets and paths and leave parts of the system under-tested [S7]. Studies of code review, dependency updates, and bug fixing report that reviewers focus on functional correctness, that certification regimes can encourage checklist compliance rather than ongoing detection, and that existing test suites often lack coverage of security-relevant scenarios [S23, S34, S45, S48, S52].

Legal and ethical concerns further limit the use of large-scale scanning or active probing. Operators of Internet-scale scanning infrastructure report uncertainty about whether benevolent scans could be interpreted as unauthorised access and worry about potential civil liability, infrastructure overload, and the difficulty of obtaining explicit consent at scale [S16]. Few tools or practices in RQ1 explicitly engage with these legal and ethical constraints, for example by providing mechanisms for consent management, opt-out, or carefully scoped probing. There is also limited support for planning review and testing coverage given fixed budgets, or for making trade-offs explicit when some SVs must remain untested.

Viewed as a whole, the alignment between RQ1 and RQ2 is uneven. Solution efforts concentrate on technical support for SV detection and SV repair, often at the level of individual tools or artefacts, while many of the most persistent challenges arise from organisational conditions (T1), adoption and integration issues (T2), and information and risk assessment problems (T3). System-level and domain-specific constraints (T4) and constraints on detection, scanning, and review (T5) are acknowledged in both primary and secondary work but are rarely the primary target of evaluated interventions. Conceptually, this pattern can be seen as a matrix where rows correspond to the RQ2 themes and columns to solution families and SVM phases. The dense cells cluster around tool-centric support for SV detection and SV repair, while cells that combine socio-technical themes with under-served phases such as SV assessment and SV disclosure remain sparse. This imbalance motivates both the practical implications we draw for current SVM practice and the research directions we outline in the following subsections.

\subsection{Practical implications and recommendations for SVM practice}
\label{sec:discussion-implications}

This subsection distils evidence-backed implications from the RQ1 and RQ2 syntheses into actionable recommendations for stakeholders who design, adopt, and sustain SVM solutions. Building on the cross-cutting synthesis in Section~\ref{sec:discussion-synthesis} and the alignment analysis in Section~\ref{sec:alignment-gaps}, the focus here is on what organisations and tool builders can do with the evidence that already exists across the 57 primary studies.

\subsubsection{Implications for organisations and teams (practitioners)}
The RQ2 themes indicate that SVM outcomes depend as much on organisational capacity and workflow design as on the technical capabilities of individual tools. Teams repeatedly report limited security training, weak prioritisation of SVM tasks, ad-hoc processes, and unclear ownership of SV triage and repair decisions, even in environments where scanning and analysis capabilities are available [S3, S4, S18, S20, S22, S23, S24, S28, S31, S37, S51, S55, S56]. These conditions suggest that improving SVM performance requires targeted investment in roles, routines, and decision structures, rather than relying on upgrades in tools alone.

Based on the practices and challenges reported across T1 to T5, we suggest the following evidence-based recommendations for organisations.

\begin{itemize}
    \item \textbf{Treat SVM as delivery-critical work with explicit ownership.} Assign clear responsibility for SV triage, prioritisation, and repair across teams and systems, particularly in large or tightly coupled environments where coordination and accountability are persistent barriers [S4, S9, S21, S39, S51].
    
    \item \textbf{Invest in role-specific and domain-specific capability building.} Provide structured training and access to security expertise for developers who engage in SV-related tasks, especially in specialised stacks such as systems programming and cryptography where generic guidance is insufficient [S4, S22, S23, S42, S56].
    
    \item \textbf{Institutionalise repeatable SVM routines.} Move from one-off scripts and informal practices toward lightweight, sustainable processes that link policies and certification needs to day-to-day development and operations [S18, S20, S28, S48, S51, S55].
    
    \item \textbf{Adopt tools with integration-first criteria.} Prioritise tools that integrate cleanly into IDE and CI/CD workflows, provide timely and interpretable feedback, and minimise tuning overhead, since poor workflow fit and noisy findings routinely erode adoption and trust [S3, S9, S14, S28, S29, S31, S34, S52, S55, S56].
    
    \item \textbf{Strengthen risk-context pipelines for triage.} Combine SV alerts with local exploitability context, dependency relationships, and system criticality rather than relying on generic severity labels or checklist-style decision rules [S17, S34, S38, S45, S57].
    
    \item \textbf{Use SBOMs and security-debt markers as operational artefacts, not compliance outputs.} Ensure SBOM generation and maintenance are tied to release and patch workflows, and interpret security-debt signals with care so they support planning without obscuring high-risk obligations [S1, S39, S54].
\end{itemize}

These recommendations imply that organisations should evaluate SVM initiatives not only by the accuracy of \textit{Detection} tools or the speed of \textit{Repair} activities, but also by whether they improve cross-phase coordination across all phases of the SVM life cycle and reduce long-lived exposure in dependency-heavy or domain-constrained systems [S10, S39, S45, S46, S57].

\subsubsection{Implications for tool builders and platform providers}
The RQ2 evidence shows that tool success is often limited by interpretability, integration, configuration burden, and trust. Even when tools are technically capable, practitioners struggle with poorly structured warnings, insufficient context for action, and fragmented security evidence across tools and work systems [S5, S11, S29, S31, S32]. This suggests that tool builders should treat usability and workflow embedding as immediate design priorities rather than secondary refinements.


We therefore recommend that tool builders and platform providers prioritise the following design and deployment considerations.

\begin{itemize}
    \item \textbf{Design for actionable feedback.} Explanations should clarify why a finding matters, what evidence supports it, and how it maps to fix options within the local code and dependency context [S29, S31, S32, S40].
    
    \item \textbf{Make integration robust and low-friction.} Tools should align with how teams already plan, review, and deploy changes, and provide stable hooks into IDE and CI/CD environments so signals arrive at the right time and place [S3, S9, S14, S28, S34, S52, S55, S56].
    
    \item \textbf{Reduce configuration cost with safer defaults.} Given that tuning requires scarce expertise, tools should ship with profiles that reflect common organisational risk postures and support gradual refinement rather than requiring heavy upfront calibration [S9, S31, S49].
    
    \item \textbf{Communicate uncertainty in AI-assisted tooling.} For ML- and LLM-based systems, interfaces should expose confidence cues and known blind spots and support safe escalation to expert review, since opacity and instability are major barriers to trust in real settings [S12, S40, S43, S49, S50].
    
    \item \textbf{Require deployment-ready evidence before broad rollout.} Before releasing SVM tools widely or promoting their broader deployment, prioritise evidence from realistic environments and consider sustained adoption, triage effort, and operational outcomes, including dependency-backlog reduction and improved patch timeliness [S21, S45, S46, S56].
\end{itemize}

Finally, a subset of studies highlights that compliance and certification regimes can unintentionally encourage checklist-oriented security work. Tool builders and standards-aligned platform providers should ensure that certification-facing reporting supports continuous SVM visibility and does not displace ongoing \textit{Detection} and \textit{Assessment} activities [S48, S17]. In parallel, designers of scanning and monitoring solutions should account for ethical and legal constraints and provide mechanisms for consent, opt-out, and scoped probing to support safer assessment practices [S16].

Together, these implications show that improving SVM in practice depends on more than deploying technically capable tools. Effective SVM requires integrated solutions that combine approaches and tools with well-supported practices, and that acknowledge organisational capacity, information quality, and domain realities as first-class determinants of success. These practical priorities also point to broader research questions, which we discuss next.

\subsection{Open research challenges and future directions}
\label{sec:discussion-future}

Beyond the practical implications discussed above, we identify five open research directions that emerge most consistently from the practitioner-centred evidence base. Building on the cross-cutting synthesis in Section~\ref{sec:discussion-synthesis} and the alignment analysis in Section~\ref{sec:alignment-gaps}, each direction is grounded in recurring gaps between the solutions reported in RQ1 and the challenges reported in RQ2, with emphasis on where practitioner needs remain only partially addressed across the SV life cycle. For brevity, we refer to these research directions as R1--R5.

\subsubsection{R1: Socio-technical SVM interventions that complement tools}
A central message of the RQ2 synthesis is that many persistent barriers to SVM are organisational rather than purely technical. Across the corpus, teams report limited security training, weak prioritisation, ad-hoc workflows, and fragile collaboration between security and development, alongside limited access to security expertise or dedicated specialists, even when they deploy sophisticated analysis or scanning capabilities [S3, S4, S18, S20, S22, S23, S24, S28, S31, S37, S51, S55, S56]. Practice-oriented contributions in RQ1.2 already offer guidelines, recommendations, and lessons learned that point toward process and communication changes, but these are rarely evaluated as structured organisational interventions in their own right. Taken together, these challenges suggest that incremental improvements in detection or repair automation will not, on their own, address the constraints that most frequently prevent SVM activities from being scheduled, sustained, or scaled. Future research should therefore explore evaluated socio-technical interventions that explicitly combine tooling with changes in roles, processes, and governance. Promising directions include role and ownership models that clarify who is accountable for SV triage and repair decisions, responding to recurring ownership ambiguities [S21, S39, S51], structured training programmes tailored to specific stacks and SV classes, given persistent gaps in domain-specific guidance and security expertise [S4, S22, S23, S42, S56], and incentive or governance mechanisms that normalise SVM work as part of delivery rather than an optional compliance add-on, addressing compliance-driven or weakly institutionalised security priorities [S18, S28, S31, S48, S51]. Studies that evaluate such interventions alongside technical tools would help bridge the current gap between tool capabilities and organisational capacity.

\subsubsection{R2: Under-studied SVM phases and cross-phase coordination}
Despite growing attention to SV detection and SV repair, practitioner evidence shows that gaps in SV assessment and SV disclosure continue to shape downstream workload and risk exposure. Teams report difficulties interpreting CVSS and severity signals and translating low-level findings into risk judgements [S17, S34, S38, S45, S57]. Still, few practitioner-centred solutions explicitly target assessment workflows or evaluate how new signals change triage decisions. The disclosure phase is even less represented in practitioner-focused secondary work. Yet practitioners describe coordination friction in disclosure and external reporting relationships, including misaligned expectations, uncertain timelines, and limited organisational support for handling incoming reports [S6, S10, S11, S20, S44]. Future research should study how weaknesses in assessment and disclosure propagate into triage backlogs, patch scheduling, and long-term exposure, reflecting evidence that information gaps and disclosure dynamics complicate repair planning in practice [S10, S39], and develop cross-phase interventions that treat SVM as an interdependent life cycle rather than a loosely connected set of tasks.

\subsubsection{R3: Evaluation in deployed workflows and over time}
A consistent theme in practitioner reports is that the success of SVM tools is shaped by workflow fit, timing of feedback, and the effort required to validate results, yet evidence remains limited on how these factors influence sustained outcomes in real deployments over time. Although all primary studies in our corpus involve industry practitioners, many evaluations of approaches and tools are short-term and small-scale, for example, task-based studies, pilot deployments in a single team, or one-off interviews about anticipated use. Across these settings, practitioners report noisy or poorly explained warnings, fragile connections to IDEs and CI/CD, and high overhead for tuning or verification, which together erode trust and reduce sustained adoption [S3, S9, S14, S28, S29, S31, S34, S40, S49, S52, S55, S56]. These findings indicate that evaluation designs that focus primarily on offline accuracy or brief user studies are insufficient to predict real-world organisational impact. Future studies should prioritise in-situ, longitudinal, and larger-scale evaluations that track sustained use, workflow integration, and outcomes such as backlog reduction or improved patch timeliness, responding to ongoing concerns about adoption and scaling in automated repair and dependency-update contexts [S21, S45, S46]. Mixed-method designs that combine task-based studies with follow-up field observations and post-deployment analyses could offer more reliable evidence of whether improvements in detection or repair translate into measurable changes in SVM practice over time.

\subsubsection{R4: Risk and information infrastructure for triage}
RQ1 highlights emerging solution lines around quantitative metrics, risk models, SBOMs, and security-debt artefacts, while RQ2 shows that the information needed for risk-based triage is often incomplete, delayed, or difficult to interpret in context. Even where metrics, exploitability models, SBOMs, or security-debt artefacts exist, practitioners report that these resources have limitations and blind spots that hinder concrete prioritisation and impact analysis [S1, S10, S17, S27, S30, S34, S38, S39, S45, S54, S57]. This gap is especially visible in dependency-heavy environments where alert volume is high but exploitability context is thin [S45, S46, S57], and in settings where generic metrics fail to capture local operational risk [S17, S27, S30]. Future research should focus on integrating these artefacts into cohesive, decision-oriented pipelines. This includes establishing stronger links between SBOM generation, dependency risk signals, and patch orchestration, addressing current limitations in SBOM completeness and workflow integration [S54, S45, S46, S57], improving the interpretability and consistency of severity and exploitability indicators, given reported problems with ambiguity and decision utility [S17, S30], and developing security-debt representations that support cross-team reporting and long-term planning without obscuring critical obligations, responding to concerns about partial or misleading risk representation [S1]. Evaluations should measure not only whether new information can be produced, but also whether it changes triage outcomes.

\subsubsection{R5: Emerging domains and Human-AI collaboration in SVM}
The practitioner evidence emphasises that domain constraints frequently define the boundary of feasible SVM strategies. Studies in systems programming, blockchain and DeFi, and dependency-heavy supply chains illustrate how immutability, deep dependency graphs, performance and resource constraints, and toolchain limitations restrict what is feasible in practice [S2, S8, S35, S36, S41, S42, S45, S46, S57]. These constraints imply that future solutions must explicitly model domain realities, such as composability risk in DeFi [S35, S36, S41] or legacy interoperability and immature analysis stacks in systems contexts [S8, S42, S56]. In parallel, ML-based and LLM-based tools amplify the need for careful human oversight. Practitioners report opacity, instability, and the need for careful human verification when using AI-assisted detection or repair [S12, S40, S43, S49, S50]. They also raise concerns about false confidence and unclear accountability when automated outputs are integrated into high-stakes workflows, especially where humans are expected to approve or deploy machine-generated fixes [S21, S40, S49, S50]. Future work should therefore investigate principled human--AI collaboration models for SVM, including interfaces that communicate uncertainty, mechanisms that support safe escalation to expert review, and governance structures that clarify responsibility for automated recommendations and repairs.

Collectively, these directions suggest that the next phase of practitioner-centred SVM research should move beyond isolated tool improvements. Progress will likely depend on integrated socio-technical interventions, stronger coverage of SV assessment and SV disclosure, evaluation designs that privilege real workflow outcomes, and information infrastructures that support risk-based decisions across diverse and evolving domains. Advancing SVM in practice will therefore require research that is not only technically innovative, but also attentive to the organisational, informational, and domain constraints that shape what practitioners can realistically adopt and sustain.

\section{Threats to validity}
\label{Sec:threats-validity}

As with any SLR, our study is subject to threats to validity. Following established SLR guidelines in software engineering \cite{kitchenham2007guidelines,zhang2011identifying,wohlin2014guidelines}, we applied several mitigation measures during search, selection, extraction, and analysis. Below, we summarise the main threats to external, internal, construct, and conclusion validity in relation to the procedures described in Section~\ref{Sec:Method}.

\subsection{External validity}
External validity concerns the extent to which our findings can be generalised beyond the set of primary studies included in this review. The search strategy (Section~\ref{Sec:search-terms}) and study selection process (Section~\ref{Sec:study-selection}) pose several threats. Although we used a multi-database search, an iteratively refined search string, and backward and forward snowballing \cite{wohlin2014guidelines}, we cannot guarantee that all relevant studies were retrieved; some may use terminology not captured by our string or may appear in other databases or grey-literature sources. In addition, our restriction to peer-reviewed English-language papers published in A* or A CORE-ranked venues may underrepresent practitioner evidence from other publication outlets, regions, or industry communities. Our requirement that studies collect primary data from industry practitioners (IC-2) also fits the goal of understanding SVM from practitioners’ perspectives, but excludes secondary-data-only studies and studies with student participants. Finally, the corpus reflects work published up to 1 July 2025, so more recent studies, especially in fast-moving areas such as AI-assisted security tools, are not captured.

\subsection{Internal validity}
Internal validity concerns potential biases in how we identified, selected, extracted, and analysed the primary studies. Study selection involved judgement when applying the inclusion and exclusion criteria, so we specified the criteria a priori (Section~\ref{Sec:inclusion-exclusion}), recorded reasons for exclusion during screening, had the first author conduct the screening, and had the second author independently verify all decisions; disagreements were resolved through discussion. Data extraction is another possible source of error, so we used a structured extraction form (Table~\ref{tab:data-extraction}), piloted it on six studies, had the second author review the pilot extraction, and discussed ambiguous cases and inconsistencies in regular team meetings during full extraction (Section~\ref{Sec:data-extraction}). For RQ2, thematic analysis of challenge-related data (D13) inevitably involved interpretive judgement, so the first and second authors independently coded challenge excerpts from six studies, reconciled differences into a shared codebook, and then reviewed new or revised codes, subthemes, and themes in weekly meetings during full coding (Section~\ref{Sec:data-analysis}) \cite{braun2006using}. We also maintained traceability from themes back to coded excerpts and source studies. Finally, the primary studies themselves may be affected by publication bias, since practitioner-centred empirical work may be more likely to report successful collaborations or solutions than failed interventions or unresolved challenges. We addressed this only partially by explicitly extracting challenge-related data (D13) and incorporating both positive and negative evidence where available.

\subsection{Construct validity}
Construct validity concerns how well our operationalisations reflect the concepts of interest, such as “software vulnerability management”, “approaches and tools”, “practices”, and “practitioner-reported challenges”. We operationalised SVM in terms of four phases (Detection, Assessment, Repair, and Disclosure) and classified each study’s area of focus accordingly (D7). Because some studies span multiple phases or use terminology that does not map cleanly to this decomposition, classification required interpretation of study goals and reported activities. We mitigated this threat by revisiting full texts when necessary, discussing ambiguous cases in team meetings, and allowing studies to be assigned to multiple phases when justified. We also distinguished between \emph{approaches and tools} and \emph{practices} based on dictionary definitions and prior SLRs in SE \cite{shahin2017continuous,dissanayake2022software}. Some contributions, such as process frameworks or socio-technical tool deployments, could reasonably fit both categories; in such cases, we classified them based on the primary emphasis reported by the authors. For RQ2, we treated challenge-related statements, explanatory text, and reported consequences as evidence of practitioner-reported challenges and captured them as qualitative excerpts in D13. This depends on how primary-study authors framed and reported practitioners’ views, so some challenges may have remained implicit or underemphasised. To reduce this threat, we read the full text of each study when extracting challenge data, not only dedicated “limitations” or “discussion” sections, and focused coding on passages where practitioners’ difficulties or concerns were explicitly described. Finally, the RQ2 themes and subthemes represent one plausible organisation of the challenge landscape. We mitigated this threat through iterative theme development, explicit naming and definitions of subthemes and themes, and traceability from each theme back to the underlying codes and excerpts (Section~\ref{Sec:data-analysis}).

\subsection{Conclusion validity}
Conclusion validity relates to the strength and scope of the inferences that we draw from the reviewed studies. 

Our synthesis is primarily qualitative and descriptive. We report counts of studies per category (e.g., by SVM phase, solution type, or theme) to characterise the evidence base, but these counts should not be interpreted as statistically representative of all SVM research or all industrial practice. In particular, some areas may be under-represented because of publication bias, limited practitioner-centred empirical work, or our inclusion restrictions, rather than because they are unimportant in practice. For RQ1, our mapping of solution types and SVM phases summarises how approaches, tools, and practices have been studied with practitioners, but it does not establish effectiveness beyond what is reported in the primary studies. For RQ2, the thematic analysis identifies recurring challenge patterns, but it cannot establish causal relationships between specific organisational or technical factors and SVM outcomes. We therefore frame our conclusions as synthesised insights and trends grounded in the available evidence rather than as strong causal claims. Finally, our interpretations may still be influenced by the authors’ backgrounds and prior experience with software security and SVM. To reduce individual bias, we discussed intermediate findings in regular team meetings over approximately six months, revisited primary studies when interpretations were uncertain, and iteratively refined the narrative in light of feedback from all coauthors. Even so, alternative interpretations remain possible.

\section{Conclusion}
\label{sec:conclusion}

This systematic literature review synthesised practitioner-centred evidence on Software Vulnerability Management (SVM) across the life cycle of software vulnerabilities (SVs). We analysed 57 primary studies that collect primary data from industry practitioners to map the solution space and the challenges reported in real-world settings.

The evidence shows an uneven research landscape. Most practitioner-focused contributions report practices such as guidelines, recommendations, lessons learned, and shared experiences, while fewer studies introduce or evaluate concrete approaches and tools. In addition, the literature remains heavily concentrated on SV detection and SV repair. Practitioner-centred work on SV assessment and SV disclosure is comparatively limited, even though weaknesses in these areas can shape downstream triage workload and long-term exposure. Our synthesis highlights that persistent barriers to SVM are not only technical. Organisational capacity, role clarity, time pressure, and security culture repeatedly influence whether SVM work is prioritised and sustained. Tooling itself also remains a double-edged sword: practitioners often face noisy findings, unclear feedback, fragile workflow integration, and costly configuration, which can erode trust and reduce long-term adoption. Beyond tools, teams continue to struggle with assembling decision-quality risk information, especially in dependency-heavy and complex architectural contexts.

Overall, this review suggests that improving SVM in practice will require more integrated interventions that combine technical capabilities with organisational support, clearer ownership, and stronger information infrastructure for risk-based prioritisation. Future practitioner-centred research should also expand coverage of assessment and disclosure and place greater emphasis on in-situ and longitudinal evaluations that capture sustained use and operational impact.

\begin{acks}
This work has been supported by the Cyber Security Cooperative Research Centre Limited whose activities are partially funded by the Australian Government’s Cooperative Research Centre Program.
\end{acks}

\bibliographystyle{ACM-Reference-Format}
\bibliography{My_References}




\end{document}